\documentclass[twocolumn,]{aastex62}

\hypersetup{linkcolor=red,citecolor=blue,filecolor=cyan,urlcolor=blue}
\newcommand{\sfr}{\ensuremath{M_\odot\,{\rm yr}^{-1}}\xspace}
\newcommand{\msun}{\ensuremath{M_\odot}\xspace}

\usepackage{graphicx,amsmath,amssymb,amstext,natbib}
\usepackage{epsf}
\usepackage{epstopdf}
\usepackage{xspace}

\graphicspath{{./}{figures/}}
\accepted{\today}
\shorttitle{An ALMA molecular gas census in two high-redshift proto-clusters.}
\shortauthors{Jorge A. Zavala et al.}
\begin{document}
\title{\sc On the Gas Content, Star Formation Efficiency, and Environmental Quenching\\ of Massive Galaxies in Proto-Clusters at $z\approx2.0-2.5$ }

\correspondingauthor{Jorge A. Zavala} \email{jzavala@utexas.edu}

\author[0000-0002-7051-1100]{J. A. Zavala}
\affil{The University of Texas at Austin, 2515 Speedway Blvd Stop C1400, Austin, TX 78712, USA}

\author[0000-0002-0930-6466]{C. M. Casey}
\affil{The University of Texas at Austin, 2515 Speedway Blvd Stop C1400, Austin, TX 78712, USA}

\author[0000-0002-0438-3323]{N. Scoville}
\affil{California Institute of Technology, MC 249-17, 1200 East California Boulevard, Pasadena, CA 91125, USA}

\author[0000-0002-6184-9097]{J. B. Champagne}
\affil{The University of Texas at Austin, 2515 Speedway Blvd Stop C1400, Austin, TX 78712, USA}

\author[0000-0001-6320-261X]{Y. Chiang}
\affil{Department of Physics \& Astronomy, Johns Hopkins University, 3400 N. Charles Street, Baltimore, MD 21218, USA}

\author[0000-0001-7147-3575]{H. Dannerbauer}
\affil{Instituto de Astrof\'isica de Canarias (IAC), E-38205 La Laguna, Tenerife, Spain}
\affil{Universidad de La Laguna, Dpto. Astrof\'isica, E-38206 La Laguna, Tenerife, Spain}

\author[0000-0003-3627-7485]{P. Drew}
\affil{The University of Texas at Austin, 2515 Speedway Blvd Stop C1400, Austin, TX 78712, USA}

\author[0000-0001-9608-6395]{H. Fu}
\affil{Department of Physics \& Astronomy, The University of Iowa, 203 Van Allen Hall, Iowa City, IA 52242, USA}

\author[00000-0003-3256-5615]{J. Spilker}
\affil{The University of Texas at Austin, 2515 Speedway Blvd Stop C1400, Austin, TX 78712, USA}

\author[0000-0001-5185-9876]{L. Spitler}
\affil{Department of Physics and Astronomy, Macquarie University, NSW 2109, Australia}

\author[0000-0001-9208-2143]{K. V. Tran}
\affil{George P. and Cynthia W. Mitchell Institute for Fundamental Physics and Astronomy, Department of Physics \& Astronomy, Texas A\&M University, College Station, TX 77843, USA}

\author[0000-0001-7568-6412]{E. Treister}
\affil{Pontificia Universidad Cat\'olica de Chile, Instituto de Astrof\'isica, Casilla 306, Santiago 22, Chile}

\author[0000-0003-3631-7176]{S. Toft}
\affil{Cosmic Dawn Center (DAWN)}
\affil{Niels Bohr Institute, University of Copenhagen, Vibenshuset, Lyngbyvej 2, DK-2100 Copenhagen, Denmark}

\begin{abstract}
% \vspace{0.4cm}
We present ALMA Band 6 ($\nu=233\,$GHz, $\lambda=1.3\,$mm) continuum observations towards 68 `normal' star-forming galaxies within two Coma-like progenitor structures at $z=2.10$ and 2.47, from which ISM masses are derived, providing the largest census of molecular gas mass in overdense environments at these redshifts. Our sample 
comprises galaxies with a stellar mass range of $1\times10^{9}\,M_\odot-4\times10^{11}\,M_\odot$ with a mean $M_\star\approx 6\times10^{10}\,M_\odot$. Combining these measurements with multiwavelength observations and SED modeling, we characterize the gas mass fraction and the star formation efficiency, and infer the impact of the environment on galaxies' evolution. Most of our detected galaxies ($\gtrsim70\%$) have star formation efficiencies and gas fractions  similar to those found for coeval field galaxies and in agreement with the field scaling relations. 
However, we do find that the proto-clusters contain an increased fraction of massive, gas-poor galaxies, with low gas fractions ($f_{\rm gas}\lesssim6-10\,\%$) and red rest-frame ultraviolet/optical colors typical of post-starburst and passive galaxies. The relatively high abundance of passive galaxies suggests an accelerated evolution of massive galaxies in proto-cluster environments. The large fraction of quenched galaxies in these overdense structures also implies that environmental quenching takes place during the early phases of cluster assembly, even before virialization. From our data, we derive  a quenching efficiency of $\epsilon_q \approx 0.45$ and an upper limit on the quenching timescale of $\tau_q<1\,$Gyr. 
%median of $\sim 2\times10^{10}\,M_\odot$
% Finally, given that these structures have not yet collapsed, our results suggest that environmental quenching galaxy pre-processing might be an important quenching mechanism in overdense structures.
% \vspace{-0.4cm}
\end{abstract}

 %galaxies: groups    ->Quitar el vspace before submission

% \vspace{0.8cm}
\section{Introduction} \label{secc:intro}

 {\it  ``Nebulae of all types except the irregular are represented among its members, but elliptical nebulae and early spirals are relatively much more numerous than among the nebulae at large. The predominance of early types is a conspicuous feature of clusters in general [...]"} \citet{Hubble1931a}.\\

It has been nearly a century since the first pieces of evidence of galaxies' properties correlating with the environment arose.
% It has been known for decades that galaxies' properties correlate with environment, as implied by the quoted reference.
In the local Universe, the higher the density of the local environment, the more likely a galaxy is to be red, massive, elliptical, and non star-forming. Galaxy clusters, some of the densest structures in the Universe, are indeed dominated by these early types  (see review by \citealt{Dressler1984a}). 

Over the past two decades, studies of clusters beyond the local Universe and up to high redshifts ($z\sim1.5$) have found a similar bimodality 
between galaxies' properties and density, such that the denser areas, corresponding to galaxy clusters, contain a higher fraction of quiescent massive elliptical galaxies than the less dense environments typical of the average field density (\citealt{Balogh1999a,Scoville2007a,Scoville2013a,Peng2010a,McGee2011a,Muzzin2012a,Papovich2012a,Newman2014a,Socolovsky2018a,Strazzullo2018a}).
Therefore, in order to understand the physical origin of this dichotomy, a problem still under debate, observations of 
clusters at earlier epochs, particularly during the earliest phases of assembly, are required. 

Proto-clusters of galaxies at high redshifts ($z\gtrsim1.5$) are hence ideal targets to study the influences of the environment in the formation and evolution of galaxies. Most of these structures, which extend over several Mpc, are known to be  undergoing an active epoch of star formation (e.g. \citealt{Geach2005a,Chapman2009a,Tran2010a,Dannerbauer2014a,Casey2016a,Miller2018a,Oteo2018a}), probing a critical transitional phase and revealing the expected `reversal' of the star formation-density relation required to explain the population of galaxies in mature clusters (e.g. \citealt{Elbaz2007a,Cooper2008a}). 

A thorough characterization of the star formation activity requires a census of the molecular gas mass, the main ingredient from which stars are formed. While some pioneering studies on the gas content of \mbox{(proto-)}cluster structures at $z\gtrsim1.5$ have been carried out, they have  
focused on extreme, rare sources like Dusty Star-Forming Galaxies (DSFGs) or Active Galactic Nucleus hosts (AGN), or on samples of a few targets (\citealt{Aravena2012a,Casasola2013a,Dannerbauer2017a,Noble2017a,Rudnick2017a,Stach2017a,Wang2018a}). 
Despite these significant efforts, the physical properties of less extreme star-forming galaxies and the impact of the environment on their formation and evolution are still far from understood.  This can only be addressed studying large statistical samples of `normal' star-forming galaxies in high density environments at different epochs.

 This study focuses on two particularly unique rich protoclusters at $z=2.10$ and  $z=2.47$
 in the COSMOS field (\citealt{Scoville2007b}). These 
structures 
extend up to half a degree on the sky, in line with the
expectations for a massive cluster
in formation, according to 
 cosmological simulations (\citealt{Chiang2013a}). This active formation phase is further support by the high number of extreme galaxies within the proto-clusters. The $z=2.10$ structure contains 9 rare 
DSFGs and  4 AGNs,
has a total star formation rate (SFR) of $\sim5300\,\sfr$, total stellar mass 
of $\sim2\times10^{12}\,M_\odot$, a galaxy overdensy of $\delta_{\rm gal}\sim8$, and an 
estimated total halo mass of $\sim2\times10^{14}\,M_\odot$ (\citealt{Spitler2012a,Yuan2014a,Hung2016a,Casey2016a}). 
Similarly, the $z=2.47$ structure contains at least 7 rare 
DSFGs and 5 AGNs, implying an overdensity of $\delta_{\rm gal}\sim10$, a
total SFR of $\sim4500\,\sfr$, total stellar mass of $\sim1\times10^{12}\,M_\odot$, and halo
mass of $\sim8\times10^{13}\,M_\odot$  (\citealt{Casey2015a,Casey2016a}). 
This proto-cluster might indeed be embedded in a larger structure including several overdensities within a redshift range of $z=2.42-2.51$
(\citealt{Chiang2015a,Diener2015a,Lee2016a,Wang2016a,Cucciati2018a,Gomez-Guijarro2019a}).
Both structures are predicted to exceed $\gtrsim1\times10^{15}\,M_\odot$ by $z=0$.
The sources targeted in this work are `normal' star-forming galaxies with confirmed spectroscopic redshifts in these two structures.
These rest-frame UV/optically selected systems are indeed expected to be more representative of the star-forming population than the extreme sources surveyed in previous studies, allowing for a detailed study on the environmental effects of star formation in a relatively large sample of 68 sources.

This paper is structured as follows: sample selection and observations are given in \S\ref{secc:S1}. In \S\ref{secc:observations}, we describe the methodology used to derive gas, stellar masses, and SFRs. The main results are presented in \S\ref{secc:analysis}, where we compare the star formation efficiency and gas content of these sources to those estimated for coeval galaxies in normal density environments. Finally, we summarize our conclusions in 
\S\ref{secc:conclusions}.
We assume a standard Planck cosmology throughout this paper, 
with $H_0 = 68\rm\,km\,s^{-1}\,Mpc^{-1}$ and $\Omega_\Lambda= 0.69$ 
(\citealt{Planck-Collaboration2016a}), and a \citet{Chabrier2003a} initial 
mass function (IMF) for SFR and $M_\star$ estimations.

\section{Observations and data analysis} \label{secc:S1}

\subsection{Target selection and control sample}\label{secc:sample}
Previous studies probing the gas content of galaxies in high density environments suffer from small samples of extreme sources like DSFGs or AGN. In contrast, our study benefits from a relatively large sample of massive galaxies which are spectroscopically-confirmed to be proto-cluster members (drawn from \citealt{Casey2015a,Casey2016a}; and \citealt{Hung2016a}). As described in detail in the aforementioned references, the targets were selected from several redshift surveys, which followed up cluster candidates from large near-infrared (NIR)-selected catalogs (mostly $K$-band selected galaxies) in the COSMOS field (e.g. \citealt{Lilly2007a,Lilly2009a,Yuan2014a,Muzzin2013a,Kriek2015a,Tasca2017a}), with the exception of a few galaxies found serendipitously through other means.

\begin{figure}[t!]\hspace{-0.6cm}
\includegraphics[width=0.53\textwidth]{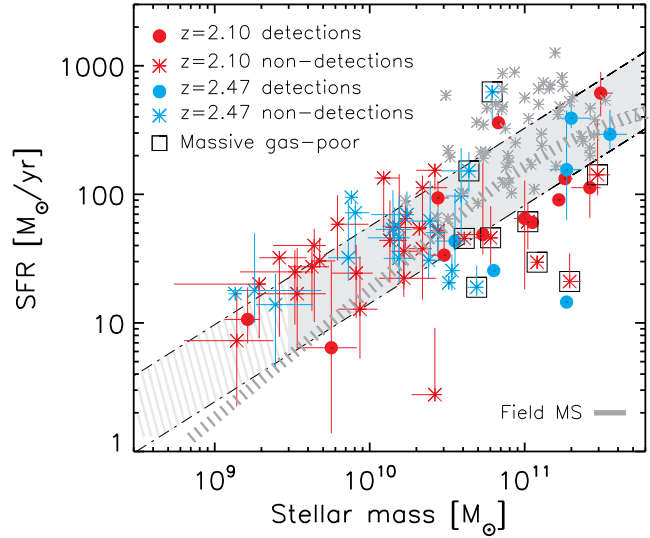}
\caption{Distribution of our targets in the SFR-M$_*$ plane in comparison with the star-forming main sequence. Sources detected with ALMA are represented by the blue and red solid circles while non-detections are identified with the blue and red asterisks. Blue symbols correspond to those galaxies in the $z=2.47$ proto-cluster while red symbols denote members of the $z=2.10$ structure. The adopted control sample drawn from \citet{Scoville2016a} is shown as gray asterisks. Additionally, two different parametrization of the star-forming main sequence at the mean redshift of our sample are shown. The gray shaded area represents the relation derived by \citet{Speagle2014a} and gray dashed line the one reported by \citet{Schreiber2015a}. Our sample spans a large range of SFR and stellar mass, representative of the main sequence population. Interestingly, at high stellar masses, an evolved population of galaxies with low SFRs seems to be manifested. Some of them (identified by the black squares) have indeed very low gas mass fractions (\S\ref{secc:gas_fraction}) and red colors (\S\ref{secc:quenching}), supporting the environmental quenching scenario discussed in \S\ref{secc:quenching}.
\label{fig:MS}}
\end{figure}

Although some proto-cluster members were probably missed in those follow-up spectroscopic campaigns, there are no obvious biases related to the sample selection 
that might affect our results. At high stellar masses our sample comprises $\sim70\%$ of the known members with a stellar mass of $M_\star\gtrsim10^{10}\,M_\odot$. The SFRs of the targets lie on (and below) the estimated star-forming main sequence (Figure \ref{fig:MS}), probing the typical star-forming galaxy population at this epoch. The fact that some of the most massive galaxies lie below the main sequence is attributed to environmental effects, as discussed in \S\ref{secc:quenching}. Note that if any bias exist such that we systematically undersample galaxies with low SFRs, it would only reinforce our conclusion about the higher fraction of passive galaxies in the proto-cluster structures. 

In summary, our final sample includes a total of 68 sources within the two structures, 27 in the $z=2.47$ overdensity and 41 in the $z=2.10$ structure. Although our work focus on galaxies with $M_\star\gtrsim10^{10}\,M_\odot$, we include several members  with $10^9\,M_\odot < M_\star < 10^{10}\,M_\odot$ which lie within the ALMA primary beam (see Tables \ref{table:catalogue1} and \ref{table:catalogue2} and Figure \ref{fig:MS}).
This is one of the largest samples of $z\gtrsim2$ galaxies in dense environments for which gas masses have been derived, 
comprising sources representative of the `normal' star-forming galaxy population, and hence, ideal to test 
our understanding of gas fueling in proto-cluster environments.

This sub-sample of massive 
sources has a good analogous control sample, which allows for a direct comparison between the properties of the field and the proto-cluster galaxies and, consequently, a better understanding of the environmental effects. The adopted field scaling relations comes from \citealt{Scoville2016a}, who  reported ALMA Band 7 ($\nu=345\,$GHz) observations towards $\sim55$ targets at the same redshifts ($\langle z\rangle=2.2$), UV-luminosities, and masses ($10^{10}\,M_\odot \lesssim M_\star <4\times10^{11}\,M_\odot$; see Figure \ref{fig:MS}), yet residing in less dense environments (\citealt{Darvish2018a}). These sources were drawn from the NIR-selected catalogs of \citet{Ilbert2013a} and \citet{Laigle2016a}, similar to those used for the identification of the proto-cluster galaxies targeted in this work. Additionally, their gas masses were derived employing exactly the same methodology adopted here (see details in \S\ref{secc:ism_mass}), and stellar masses and star formation rates were also estimated following similar procedures (see \S\ref{secc:sed_fitting}), allowing us to make a balanced comparison.

% It should be noted that if we instead use the scaling relations derived from different works (e.g. \citealt{Tacconi2018a}), the general conclusions of the paper remain the same. 

Given that the completeness of our sample drops at $M_\star\lesssim10^{10}\,M_\odot$ where, additionally, we lack a control sample, the conclusions from this work focus only on the most massive objects with $M_\star\gtrsim10^{10}\,M_\odot$.

\subsection{ALMA observations and detection of dust continuum}\label{secc:data}
ALMA Band 6 observations were conducted on 4 April 2017 as part of the Cycle 4 program 2016.1.00646.S (PI: C. Casey), using the 12\,m antennae in a relatively compact configuration (with the longest baselines at 0.46\,km). These observations comprise a total of 46 pointings with an average on-source integration time  
of $\sim5$\,min, encompassing a total of 68 spectroscopically-confirmed galaxies within the two studied proto-cluster structures. The correlators were configured to maximize the bandwidth in order to increase the continuum sensitivity, providing a total bandwidth of 7.5\,GHz centered at 233\,GHz ($\approx1.3$\,mm). 

Data reduction was individually performed following the ALMA reduction pipeline scripts in CASA (version 4.7.2). A few noisy channels in two different spectral windows were flagged for all the pointings before imaging. The continuum maps were obtained using a natural weighting of the visibilities in order to obtain the highest sensitivity, and using a pixel size of $0.12''$ (roughly seven times smaller than the beam size), yielding central noises between $\sigma_{1.3\rm mm}=40-50\,\rm\mu Jy/beam$ and a typical synthesized beamsize of $\theta_{\rm FWHM}\approx0.9''$. 
% 

% \subsection{Dust continuum emission and ISM masses}
These continuum images are used to search for dust emission for each individual galaxy within $1''$ of their respective optical positions. This search radius accounts for any astrometry offset between the ALMA and the optical images (e.g. \citealt{Dunlop2017a}) and for real misalignments between the dusty regions and the bulk of the stellar emission (e.g. \citealt{Swinbank2010a}; \citealt{Hodge2016a}). To derive the flux densities a simple peak-finding algorithm is implemented on the primary-beam corrected images. A source is considered detected only if it satisfies a detection threshold of $\ge3\sigma$, for which less than one spurious detection is expected given our positional priors. In this process, the local noise is computed to be the $68^{th}$ percentile of the distribution of pixel values, which corresponds to $\pm1\sigma$ for a  Gaussian distribution. The errors calculated with this method are consistent with those derived by measuring the standard deviation of small off-source apertures. From all the 68 proto-cluster galaxies which lie within our surveyed area, 19 sources were detected above our adopted threshold, implying a detection rate of $\approx 25-30\%$ in each structure. The derived flux density of the detected  targets, and the $3\sigma$  upper limits of the non-detections,  are reported in Tables \ref{table:catalogue1} and \ref{table:catalogue2}, while HST/ALMA cutouts are shown in Appendix \ref{appendix2}.

To derive better constraints on the average flux densities of the non-detections, a stacking analysis was performed. We extract small cutouts in the image plane centered at the position of each galaxy and then combine them in a weighted average. Weights are estimated as the squared inverse of the noise measured around each source. Finally, the adopted stacked flux density (or upper limit if $<3\sigma$) corresponds to the maximum value within $1''$ of the center of the stacked image. This stacking procedure was performed for different sub-samples defined by SFR, stellar mass, or redshift, which are reported in Table \ref{table:catalogue3}.

% The local noise is estimated to be the $68^{th}$ percentile of the distribution of pixel values from a region of $6''\times6''$ centered at the position of the source

\section{Derivation of gas masses, stellar masses, and star formation rates.} \label{secc:observations}

\subsection{ISM masses}\label{secc:ism_mass}
\citet{Scoville2014a,Scoville2016a} developed the physical and empirical bases for using the long wavelength Rayleigh-Jeans  dust emission  as a probe of the ISM gas content of galaxies, calibrating a ratio between the specific luminosity at rest-frame $850\,\rm\mu m$ to the total ISM mass. Our ALMA observations trace the rest-frame $\sim 410$ and $350\,\rm\mu m$ emission for the $z\sim2.10$ and 2.47 proto-cluster galaxies, respectively, probing hence the Rayleigh-Jeans regime. Following \citealt{Scoville2016a} we derive the ISM masses using:
  \begin{eqnarray}
M_{\rm mol}   &=& 1.78 ~S_{\nu_{\rm obs}}[\rm mJy]  ~ (1+z)^{-4.8} ~   \left({ \nu_{850\mu \rm m}\over{\nu_{obs} }}\right)^{3.8} \times  \nonumber  \\
 && ({\it{D}_L \rm{[Gpc ]}})^{2}  \left\{{6.7\times10^{19}\over{\alpha_{850\rm\mu m}  }} \right\}    ~{{\Gamma_{0}} \over{{\Gamma_{RJ}}}}~   ~10^{10}\,\rm M_\odot, 
  \end{eqnarray}
where $\alpha_{850\rm\mu m}$ is the empirically callibrated ratio between long-wavelength dust luminosity and molecular gas (i.e. $\alpha_{850\rm\mu m}\equiv\langle L_{\nu_{850\rm\mu m}}/M_{\rm mol}\rangle$), $\nu_{\rm obs}=233$\,GHz is the observed frequency, $\it{D}_L$ is the luminosity distance at the redshift $z$, and $\it{\Gamma_{RJ}}$ is the correction for the departure in the rest frame of the Planck function from Rayleigh-Jeans (i.e. $B_{\nu_{\rm rest}}/\rm RJ_{\nu_{\rm rest}}$) given by
\begin{eqnarray}
{\Gamma}_{\rm RJ}(T_{\rm d},\nu_{\rm obs}, z) &=&  {h \nu_{\rm obs} (1+z) / k T_{\rm d} \over{e^{h \nu_{\rm obs} (1+z) / k T_{\rm d}} -1 }} ~,
  \end{eqnarray}
with   ${ \Gamma_{0}} = {\Gamma}(T_{\rm d}=25\,{\rm K}, \lambda=850\,{\rm\mu m}, z=0)$. The uncertainties in the derived masses from this callibration are expected to be less than 25\% (\citealt{Scoville2016a}).

ISM masses were calculated for all the ALMA detected galaxies and their associated errors were obtained by propagating the uncertainties on the flux densities. For those galaxies not detected in continuum emission, $3\sigma$ upper limits on the ISM were derived using the flux density upper limits. Additionally, the stacked fluxes described above were also used to derive stacked ISM masses (or upper limits in case of non-detections) for different sub-samples divided by SFR or stellar mass. These results are summarized in Tables \ref{table:catalogue1}, \ref{table:catalogue2}, and \ref{table:catalogue3}.

\subsection{Stellar masses and star formation rates}\label{secc:sed_fitting}
The large ancillary data available in the COSMOS field provide an exquisite set of photometric measurements in multiple bands for our sample of galaxies, which can be used to estimate stellar masses and star formation rates through Spectral Energy Distribution (SED) modeling. We use the COSMOS2015 catalog (\citealt{Laigle2016a}; the same used by \citealt{Scoville2016a}, from which our control sample is extracted) to obtain the photometry of each of our targets by matching counterparts within a $1''$ radius. The catalog includes {\it YJHK$_s$} observations from the UltraVISTA survey, Y-band images from the Subaru telescope, infrared photometry from {\it Spitzer}, in addition to other data spanning from {\it GALEX} near ultraviolet to SPIRE/{\it Herschel} far-infrared, comprising more than 30 bands. 

We perform our own SED fitting using all the 
available photometry, including our new ALMA data, 
using the {\sc MAGPHYS} code (\citealt{da-Cunha2008a,da-Cunha2015a}), which adopts an energy balance technique between the stellar and dust emission. 
This approach implies that the stellar component is coupled to the dust-emitting region.
While this does not necessarily apply for extreme DSFGs (e.g. \citealt{Hodge2016a}), it usually does for less extreme star-forming galaxies like those studied in this work.
The fitting was done with a fixed redshift according to the spectroscopic redshift of each source, using the spectral population synthesis models of \citet{Bruzual2003a}, with a \citet{Chabrier2003a} IMF, and  a continuous delayed exponential star formation history, similar to the parameters adopted in the work by \citet{Laigle2016a}. The stellar masses derived for the galaxies presented here are in good agreement with those reported in the COSMOS2015 catalog with a mean ratio of ${\rm log}(M_{*_{\rm \small{COSMOS}}})/{\rm log}(M_{*_{\rm  \small{MAGPHYS}}})=1.01\pm0.06$. All the estimated SFRs and $M_\star$ are also reported in Tables \ref{table:catalogue1} and \ref{table:catalogue2} and shown in Figure \ref{fig:MS}.

\section{Results} \label{secc:analysis}
\subsection{The star formation efficiency}\label{secc:SFE}

The star formation efficiency, typically measured as the SFR per unit gas mass (${\rm SFE\equiv SFR}/M_{\rm mol}$), is an important quantity in the understanding of the star formation activity and stellar mass growth, which is directly linked to the gas depletion timescale ($\tau=1/\rm SFE$). Testing its universality or dependence on other parameters, such as stellar mass, redshift, or environment, is required in order to have a complete view of the structure formation in the Universe. 

Our observations and the measurements of the molecular gas content and SFR described above (see \S\ref{secc:observations}) allow us to study the SFE in one of the largest and most complete samples of   proto-cluster galaxies at $z>2$, sheding light on the influence of the environment on this quantity.

\begin{figure}[t!]\hspace{-0.6cm}
\includegraphics[width=0.53\textwidth]{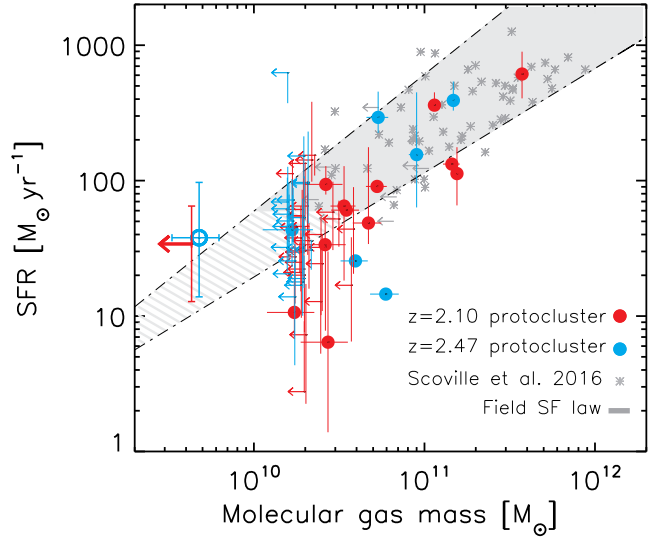}
\caption{The SFR-$M_{\rm mol}$ relation as a proxy for the SFE. The $z=2.10$ and 2.47 proto-clusters member galaxies detected by ALMA are represented by the red and blue filled circles, respectively, while the individual non-detections are plotted as $3\sigma$ upper limits (small red and blue left arrows). The $3.3\sigma$ detection from the stacking of the non-detected galaxies with $\rm SFRs=10-100$ in the $z=2.47$ protocluster is illustrated by the large open blue circle, while the $3\sigma$ upper limit derived from the stacking of the analogous galaxies in the $z=2.10$ proto-cluster is illustrated by the large red left arrow. Most of our detections and upper limits are consistent with the star formation law found for field galaxies at similar redshift, represented by the gray shaded area and individual gray asterisks (\citealt{Scoville2016a}). Those sources lying below the relation and those whose upper limits lie above it are discussed in the main text.
\label{fig:SFE}}
\end{figure}

In Figure \ref{fig:SFE} we explore the SFE of these proto-cluster galaxies via the SFR-$M_{\rm mol}$ plane. Thirteen out of the 18 detected galaxies shown in this plot ($\sim70\%$) lie, within the error bars, on the field scaling relation. 
To quantitatively compare the SFEs of the proto-cluster galaxies with those of our control sample, we perform a Kolmogorov-Smirnov test between the two distributions, from which we derive a probability of $p=0.81$ that both of them are drawn from the same parent distribution (when including only those detected galaxies with $M_*>10^{10}\,\msun$, if we include all the detections the probability is $p=0.27$). 

At the probed evolutionary stage of these proto-clusters,  the SFEs of the most massive galaxies show consistent values to those expected for coeval field galaxies. Although a non-negligible fraction  of the detected galaxies seems to lie below the field relation while some upper limits suggest galaxies lying above it. From the four detected sources that lie significantly below  the field relation (or to the right of it), two are part of the $z=2.10$ proto-cluster and two from the $z=2.47$ structure. The two sources from the $z=2.47$ proto-cluster also lie below the star-forming main sequence (Figure \ref{fig:MS}), which implies that their low SFE is driven by the reduced SFR rather than by an excess of molecular gas. Indeed, these source lie on the expected relation if their offset from the main sequence is taken into account, as shown in  \S\ref{appendix0}. The two sources from the $z=2.10$ structure show SFRs in agreement with main-sequence galaxies and their high SFEs seem to be caused by their enhanced molecular gas masses (see \S\ref{appendix0}). Interestingly, 
most of the results derived from the stacking of the non-detections, which probe -- on average -- galaxies with 
 $M_*\sim 1\times10^{10}\,\msun$ (see Table \ref{table:catalogue3}), seem also to be in agreement with the extrapolation of the  field scaling relation with a few sources lying above it (see \S\ref{appendix0}).

% The rest of the detected sources show, systematically, lower SFEs, particularly, four sources that depart significantly from the field relation. Two from the $z=2.10$ proto-cluster and two from the $z=2.47$. 
% The two sources from the $z=2.47$ proto-cluster lie also below the star-forming main sequence (Figure \ref{fig:MS}), which implies that their low SFE is driven by the reduced SFR rather than by an excess of molecular gas. These source might indeed lie on the expected relation if their offset from the main sequence is taking into account, as shown in  \S\ref{appendix0} (Figure \ref{fig:SFE_msoffset}). On the other hand, the two sources from the $z=2.10$ structure show SFRs 
% in agreement with main-sequence galaxies, which might suggest that they have enhanced molecular gas masses. Nevertheless, these sources have lower stellar masses ($\sim 2\times10^{9}\,\msun$ and $\sim 6\times10^{9}\,\msun$, respectively; see Figure \ref{fig:MS}) than the rest of the detected galaxies and, as mentioned above, the low completeness of the sample at this stellar mass range and the low number of detections prevent us from deriving strong conclusions for these less massive galaxies. Despite this, the results derived from the stacking of the non-detections, which probe - in average - galaxies with 
%  $M_*\sim 1\times10^{10}\,\msun$ (see Table \ref{table:catalogue3}), seems also to be in agreement with the extrapolation of the  field scaling relation.  

Heterogeneous results regarding the SFEs of proto-cluster galaxies have been presented in the literature, including some in agreement with our results. For example, \citet{Lee2017a} detected $\rm CO(3-2)$ line emission in seven star-forming galaxies associated with the proto-cluster 4C23.56 at $z=2.49$ and derived a median SFE consistent with the reference sample (although different results have been reported in the same structure, as discussed below; \citealt{Tadaki2019a}). \citet{Dannerbauer2017a} detected an extended $CO(1-0)$ emitting disk in a proto-cluster member galaxy at $z\approx2.15$, whose gas properties and SFR follow the same relation as normal field galaxies.  Similarly, \citet{Darvish2018a} used ALMA dust continuum  observations to investigate the role of environmental density on the molecular gas content in a large sample of $\sim400$ massive  ($M_\star\gtrsim10^{10}\,M_\odot$) galaxies within $z\approx0.5-3.5$, where the density  was estimated by the projected surface density of galaxies over different redshift slices, and concluded that the SFE is independent of galaxy overdensity. \citet{Wang2018a} obtained $\rm CO(1-0)$ observations towards a concentrated group of 14 galaxies at $z\approx2.51$ which might be associated with the $z\approx2.47$ proto-cluster structure presented here (see also \citealt{Cucciati2018a}; Champagne et al. in prep.). Their estimated SFEs show a large variety of values, although an interesting tentative trend suggests a high SFE towards the center of the small surveyed core. \citet{Gomez-Guijarro2019a} re-analyzed the \citeauthor{Wang2018a} sample with deeper observations and present new CO observations in two new proto-clusters at $z=2.13$ and 2.60, finding rather similar SFEs to the field. Additionally, studies of more evolved (collapsed) clusters at $z\sim1.6-2.1$ targetting CO emission lines have found similar SFEs or depletion timescales ($\tau=1/\rm SFE$) to the ones measured in field galaxies at similar redshifts (\citealt{Rudnick2017a}).

On the other hand, \citet{Tadaki2019a} presented $\rm CO(3-2)$ observations towards 66 H$\alpha$-selected galaxies in three proto-clusters at $z=2.16$, 2.49, and 2.53. Interestingly, in the stellar mass range of $10.5<log(M_\star/M_\odot)<11.0$ they reported SFEs lower (i.e. longer depletion timescales) than expected from the scaling relation, although these galaxies only represent $30\%$ of all the sources within this stellar mass range. The rest of galaxies (including all those with $M_\star>10^{11}\,M_\odot$) are in agreement with the field relations.
In line with these results, \citet{Noble2017a} found depletion timescales systematically higher than the scaling relation for nine $\rm CO(3-2)$-detected galaxies in a cluster at $z\sim1.6$, although most are within a one standard deviation of the relation (plus several non-detection whose upper limits are in agreement with the field). Similarly, \citet{Hayashi2018a} presented $\rm CO(3-2)$ observations in several cluster galaxies at $z\sim1.5$ finding again systematically longer depletion timescales, although $\sim50\%$ of their sample might be in agreement the expected relation for the field (when non-detections are taken into account).

Although our results suggest that most of the massive ($M_\star>10^{10}\,M_\odot$) galaxies have SFEs similar to coeval field galaxies, 
it is clear that other works in the literature have revealed  (proto-)cluster galaxies with a large variety of properties, with some of them having SFEs similar to coeval field galaxies  and  others lying off the expected relations. Larger samples of high-redshift clusters and proto-clusters, as well as deeper observations to mitigate non-detections, are hence required to fully understand the star-formation activity and gas fueling in these overdense structures.

% Tadaki: 3/7 lie above the scaling relation (only 2/7 if errors are taken into account) +3 non-detections (so onlt 3/10 or 2/10 ~30\%)

%c.f. Hayashi+18 no strong dependence of gas fraction and depletion time on the clustercentric radius and accretion phase

\subsection{Gas content and gas fraction}\label{secc:gas_fraction}

\begin{figure}[t!]\hspace{-0.6cm}
\includegraphics[width=0.53\textwidth]{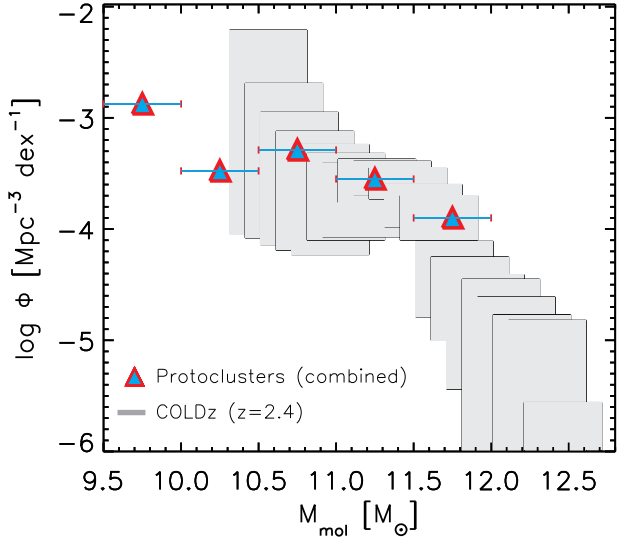}
\caption{Comparison between the COLDz gas mass function derived from $\rm CO(1-0)$ observations at $z\approx2.4$ (gray squares, \citealt{Riechers2018a}) and the gas mass function derived from our ALMA follow-up of proto-cluster galaxies at similar redshifts (colored triangles). As described in the text, our measurements are formally lower limts since only a fraction of the proto-cluster members were observed. Therefore, it is likely that, after taken into account the incompleteness effects,  the proto-clusters show an enhanced gas mass function (and hence gas volume density) compared to the field.
\label{fig:gas_lum}}
\end{figure}

Figure \ref{fig:gas_lum} shows the combined gas mass function of the two proto-clusters, which is formally a lower limit given the incompleteness of our follow-up survey (only a fraction of the known proto-cluster member were obseved; see \S\ref{secc:sample}) and the possible existence of more unidentified member galaxies. This estimation  assumes  a volume of $15000\,\rm cMpc^3$ for each structure, as estimated by \citet{Casey2016a}. Note that this volume encompasses the $z\approx2.47$ overdensity studied here, but not the larger structure reported in the literature with a redshift range of $z=2.42-2.51$ (\citealt{Cucciati2018a}; although the volume would only increase by a factor of $\sim2$). The first bin of our gas mass function comes from the stacked detection of 20 galaxies (14 in the $z=2.10$ structure and 6 in the 2.47; see Table \ref{table:catalogue3}), while the rest come directly from the individual detections.

The derived lower limit of the proto-clusters gas mass function is of the same order of magnitude as the field gas mass function at these redshifts\footnote{To convert the COLDz $\rm CO(1-0)$ luminosity function reported by \citet{Riechers2018a} to gas mass function, we adopt a conversion factor of $\alpha_{\rm CO}=6.5\,M_\odot\rm\,(K\,km\,s^{-1}\,pc^2)^{-1}$ as used by \citet{Scoville2016a} while calibrating the dust continuum-gas mass relation adopted in this work.}, as shown in  Figure \ref{fig:gas_lum}). Therefore, it is likely that these proto-clusters have an enhanced gas mass function or a higher gas volume density than the one measured in the field, after taking into account all the incompleteness described above. Indeed, enhanced gas volume densities have been  measured for other similar overdense structures (e.g. \citealt{Lee2017a}).

\begin{figure}[t!]\hspace{-0.6cm}
\includegraphics[width=0.53\textwidth]{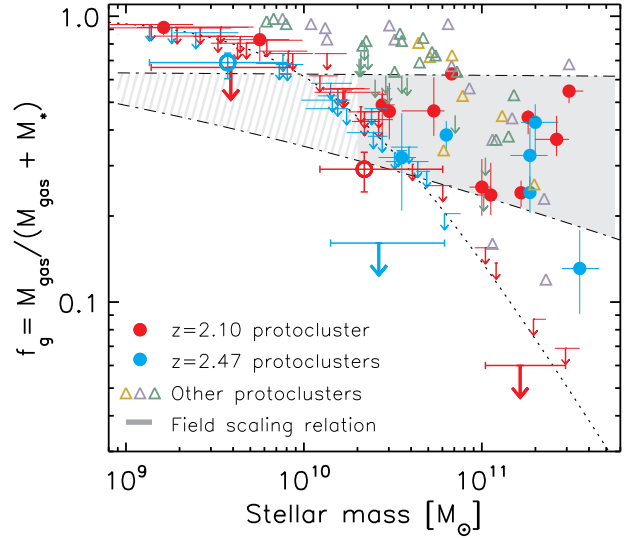}
\caption{The molecular gas fraction of the proto-cluster members as a function of stellar mass (blue and red for the $z=2.10$ and 2.47 structures, respectively), along with other measurements from the literature. 
Solid circles represent the ALMA detected galaxies while the small downward arrows are the respective upper limits for the individual non-detections. Large open circles and large downward arrows represent the results from the stacking of the non-detections of subsamples divided by stellar mass and redshift (see Table \ref{table:catalogue3}). The typical $3\sigma$ detection limit of our survey is illustrated by the dotted line.
The derived gas mas fraction of most of the detected galaxies are in  good agreement with those measured for field galaxies at similar redshifts and with the field scaling relation (gray shaded region, \citealt{Scoville2016a}), implying that these sources do not show enhanced gas masses. 
For comparison, results from studies on other proto-clusters are also included (gold, gray, and green triangles for \citealt{Lee2017a}, \citealt{Gomez-Guijarro2019a}, and \citealt{Tadaki2019a}, respectively), showing both galaxies with enhanced gas fractions and galaxies in agreement with the field (in addition to several non-detections). Interestingly, in the systems studied in this work, there are several massive non-detected galaxies for which the estimated stacked upper limits on their gas mass lie significantly below the expected field relation (large blue and red downward arrows), resembling those measured for passive quiescent galaxies and suggesting that they are likely transitional galaxies soon-to-be quenched (see discussion in \S\ref{secc:quenching}). 
\label{fig:gas_fract}}
\end{figure}

In Figure \ref{fig:gas_fract} we plot the derived gas mass fraction, $f_{\rm gas}=M_{\rm mol} / (M_{\rm mol} + M_\star)$, as a function of stellar mass for the proto-cluster galaxies, and other samples taken from the literature, including our reference sample. As it can be seen, most of the ALMA-detected  galaxies, represented by the blue and red filled circles in the figure, show gas fractions that resemble those estimated for coeval field galaxies and their scaling relation\footnote{The scaling relation plotted in Figure \ref{fig:gas_fract} was estimated via the $\rm SFR-M_{mol}$ relation from \citealt{Scoville2016a}, transforming the SFR to stellar mass using the main sequence relation reported by \citealt{Speagle2014a}. }, which is illustrated by the gray shaded region (see also \S\ref{appendix0}). 
The individual upper limits derived from the ALMA non-detections for those galaxies with $M_\star\lesssim5\times10^{10}\,M_\odot$ are also consistent with the field relation, but those of the most massive galaxies point towards reduced gas fractions when compared to the field (see also \S\ref{appendix0}).

Different works from the literature have also reported gas fractions in proto-cluster galaxies in agreement with the field. For example,  \citet{Lee2017a} found that most of the observed galaxies in a $z=2.49$ proto-cluster show gas fractions comparable to those of field galaxies. Their detections are plotted in Figure \ref{fig:gas_fract} (gold triangles), where it can be seen that their values are in agreement with our measurements.
% Only 3 out of the 22 targeted galaxies lie above the field relation, showing an excess of gas fraction (those with only upper limits are not included in the figure). 

 Figure \ref{fig:gas_fract} also shows the results from \citet{Tadaki2019a} and \citet{Gomez-Guijarro2019a}  who observed galaxies in several overdensities within $z=2.17-2.60$. Beside the longer gas depletion timescales found by \citet{Tadaki2019a} for some galaxies with $10.5<log(M_\star/M_\odot)<11.0$, 
 they also found sources with high gas mass fractions (although they represent $\lesssim50\%$ of all the  galaxies within this stellar mass range). Most of the remaing sources, and all the galaxies with $M_\star\gtrsim10^{10}\,M_\odot$, have gas fractions (or upper limits) in agreement with the field relation. Similarly, \citet{Gomez-Guijarro2019a} found some galaxies with large gas fractions (all of them with $M_\star\lesssim6\times10^{10}\,M_\odot$) despite  showing SFEs consistent with the field. 
 Nevertheless, most of these gas-rich galaxies lie above the main sequence, and therefore, might be more representative of the starburst population. 
 Galaxies with enhanced gas fractions have also been reported in more evolved (collapsed) cluster at lower redshifts, like those reported by \citet{Noble2017a,Hayashi2018a} at $z\sim1.5$.

 While some proto-clusters show most of their members with gas fractions in agreement with coeval field galaxies, as those presented in this work (see also \citealt{Lee2017a}), there is clear evidence that other structures have individual systems with large gas masses (e.g. \citealt{Tadaki2019a,Gomez-Guijarro2019a}), as discussed above. Discriminating between if these gas-rich systems exist preferentially in overdense environments or if they represent only outliers of the general population (note that this kind of outliers also exists in the field)  requires  deeper observations  of complete samples, even deeper than the ones presented here. For example, the results of \citet{Noble2017a}, who presents one of the best lines of evidence for the existence of sources with enhanced gas masses,  are based on the detection of only seven indivudal sources from a parent sample of 49 cluster members. Similarly, the gas-rich galaxies found by \citet{Hayashi2018a} come from 18 detections out of a parent sample of 65 sources. Although cluster-to-cluster variations, reflecting different evolutionary stages (e.g. \citealt{Shimakawa2018a,Gomez-Guijarro2019a}) are expected, it is still unclear if the bulk of the population in the aforementioned cluster also have enhanced gas fractions. Unfortunately, deeper ALMA spectroscopic observations would require several hours of on-source time per target, making spectroscopic observations of larger samples prohibitive. Continuum observations as a proxy for gas masses (\citealt{Scoville2016a}) will therefore play a determinant role in our understanding of the star formation activity during the early phases of clusters assembly.

\subsection{Environmental quenching}\label{secc:quenching}

 As shown in Figure \ref{fig:gas_fract}, some of the most massive galaxies studied in this work show very low gas mass fractions, $f_{\rm gas}\lesssim 6-10\,\%$ (see also \citealt{Wang2018a}), resembling those estimated for quiescent galaxies (note, however, that most of the constraints on the gas the gas fraction of passive galaxies are limited to $z<2$;  e.g. \citealt{Sargent2015a,Gobat2018a,Spilker2018a,Bezanson2019a}). These massive gas-poor galaxies might be in a transitional phase
towards a quiescent mode, and hence, they are ideal 
sources to study the quenching mechanisms, especially 
the effects of environmental quenching.

In Figure \ref{fig:MS} we highlight the most massive gas-poor systems ($M_*>4\times10^{10}\,M_\odot$) with black squares. As it can be seen, most of them (7 out of 9) lie below the star-forming main sequence, supporting a possible quenching phase. The galaxy that lies above the main sequence is known to host an AGN (based on an X-ray detection; \citealt{Laigle2016a}), and therefore, it is likely that its SFR might be overestimated (note that only six of the whole sample are classified as AGN, as mentioned below).

\begin{figure}[t!]\hspace{-0.6cm}
\includegraphics[width=0.53\textwidth]{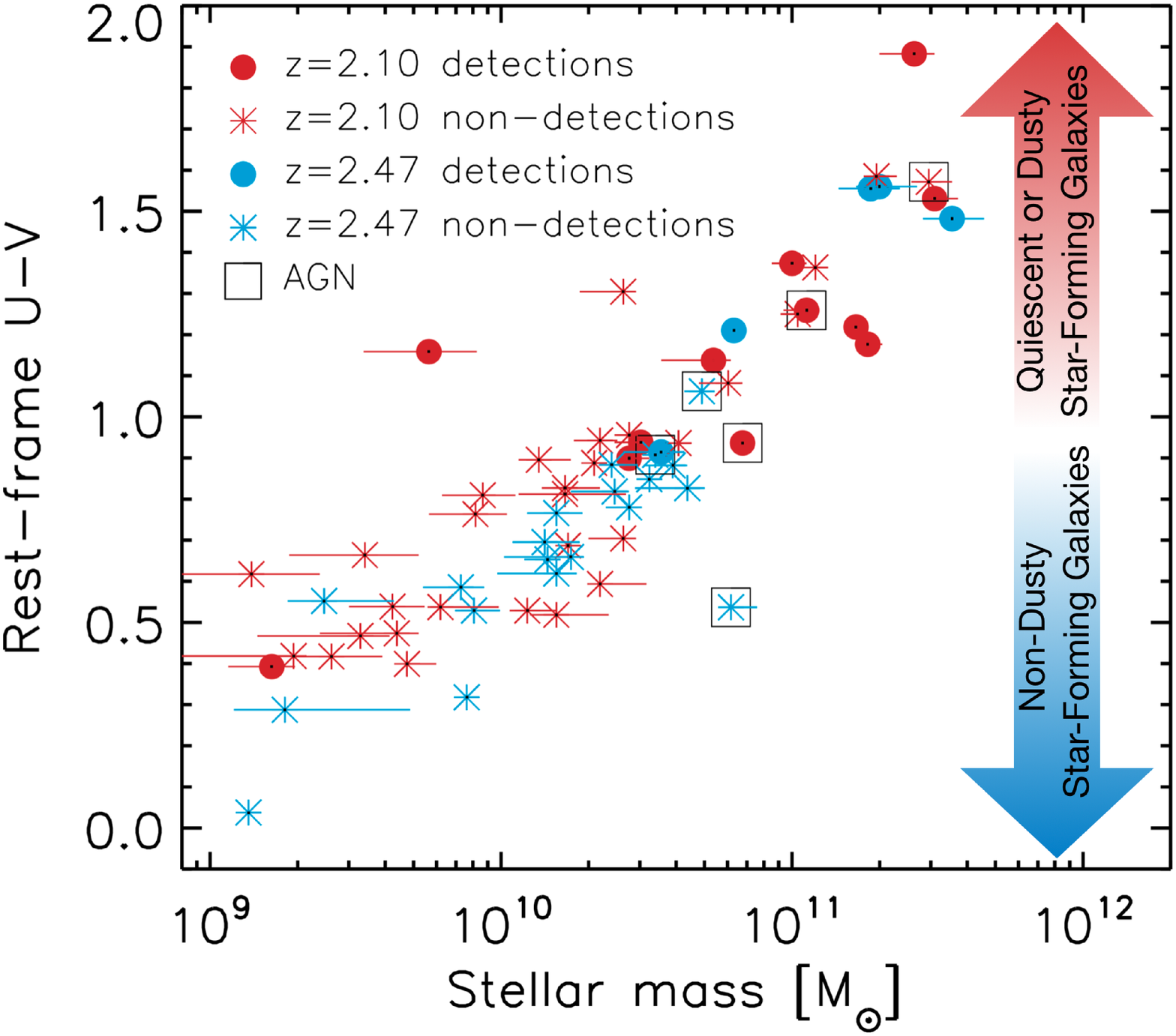}
\caption{The rest-frame U-V color-stellar mass relation. Galaxies detected with ALMA are represented by the filled blue and red solid circles (for those in the $z=2.47$ and 2.10 protoclusters, respectively), while the asterisks represents the non-detections. The most massive galaxies show redder colors than the less massive and include both dusty star-forming sources and quiescent evolved gas-poor systems. Those galaxies hosting an AGN -identified via X-ray detection- are indicated with a black open square.)
\label{fig:UVcolors}}
\end{figure}

To further explore their possible passive nature,  Figure \ref{fig:UVcolors} shows 
the rest-frame U-V color of our sample of proto-cluster galaxies, which has been proven to be a good indicator to disentangle the quiescent and star-forming populations of galaxies via  blue and  red populations (e.g. \citealt{Bell2004a,Brown2007a,Faber2007a}). This figure clearly shows that  
the most massive galaxies undetected by ALMA have red colors, as it would be expected if they are in the quenching process. 

\begin{figure}[t!]\hspace{-0.6cm}
\includegraphics[width=0.53\textwidth]{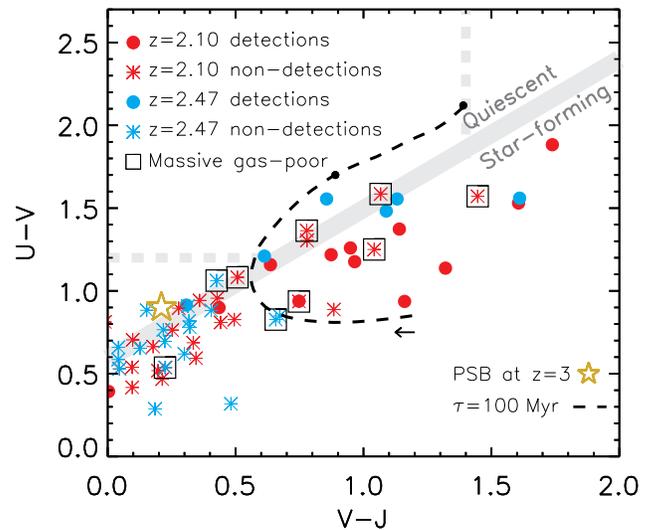}
\caption{ Rest-frame $U-V$ vs. $V-J$ color-color diagram (not corrected for dust attenuation). Proto-cluster galaxies within the $z=2.41$ structure are represented by the blue symbols while those in the $z=2.10$ are shown in red. Solid cirlces represent ALMA-detections while asterisks indicate non-detections.  The cut proposed by \citet{Belli2019a} to separate star-forming and quiescent galaxies is indicated by the gray diagonal solid line (dashed lines indicates additional constraints, e.g. \citealt{Muzzin2013a}). The most massive ($M_\star>4\times10^{10}\,M_\odot$) gas-poor galaxies in our sample (illustrated with black squares) are thought to be post-starburst or transition galaxies (see discussion in \S\ref{secc:quenching}). As reference, a well-studied post-starburst galaxy at $z\approx3$ (\citealt{Marsan2015a}) is plotted as a gold star as well as the predicted color evolution for a fast quenching path described by a tau model with a short timescale of $\tau=100\,$Myr (\citealt{Belli2019a}). 
\label{fig:UVJ}}
\end{figure}

In Figure \ref{fig:UVJ} we show instead the rest-frame (U-V) vs (V-J) color-color diagram. This UVJ plot is commonly used to distinguish between dusty, 
star-forming galaxies and quenched galaxies since both populations have usually red colors (e.g. \citealt{Wuyts2007a,Williams2009a}). While it is true that the massive gas-poor galaxies described above 
are not entirely within the quiescent parameter space, they lie very close to it. Additionally, it as been shown that some passive galaxies remain outside of the  UVJ selection area well after their actual quenching (up to almost $0.5\,$Gyr), particularly, for those in which the SFR decrases abruptly, called post-starbust galaxies (\citealt{Merlin2018a,Belli2019a}). To illustrate this, we plot in Figure \ref{fig:UVJ} the expected path on the UVJ diagram of a galaxy with a fast quenching process described by a tau model with a short timescale of $\tau=100\,$Myr (see details in \citealt{Belli2019a}). The colors predicted by this scenario (see also \citealt{Merlin2018a} for similar results) are in agreement with those measured for some of the massive gas-poor galaxies in our sample. The well-studied post-starburst galaxy at $z\approx3$ reported by \citet{Marsan2015a}, has, actually, very similar colors (see Figure \ref{fig:UVJ}). 

These findings, combined with the ALMA non-detections, suggest that the massive gas-poor sources shown in Figure \ref{fig:gas_fract} are 
evolved galaxies, most likely post-starburst or in transition to a quiescent mode.

% ABOUT THE UVJ FILTERS (from Wild et al.2014)
% for each galaxy in our sample we perform a traditional K-correction by identifying the best-fit stochastic burst model SED, and then measuring the rest-frame U- V and V-J colours from the best-fit model5. To reproduce previous
% results from the literature we use the Bessell (1990) U and V filters, and the UKIRT J filter which is on the Mauna Kea system (Toku- naga et al. 2002). -> I did not use that 

Is the evolution of these high-mass gas-poor transition galaxies driven by the overdense proto-cluster environment? Quiescent massive galaxies have been found in the field up to $z\gtrsim3$ (e.g. \citealt{Straatman2014a,Glazebrook2017a}), and therefore, a comparison between the field and these proto-cluster galaxies is necessary to understand the impact of the environment on the quenching process. 

\begin{figure*}[t!]
\begin{center}
\includegraphics[width=0.7\textwidth]{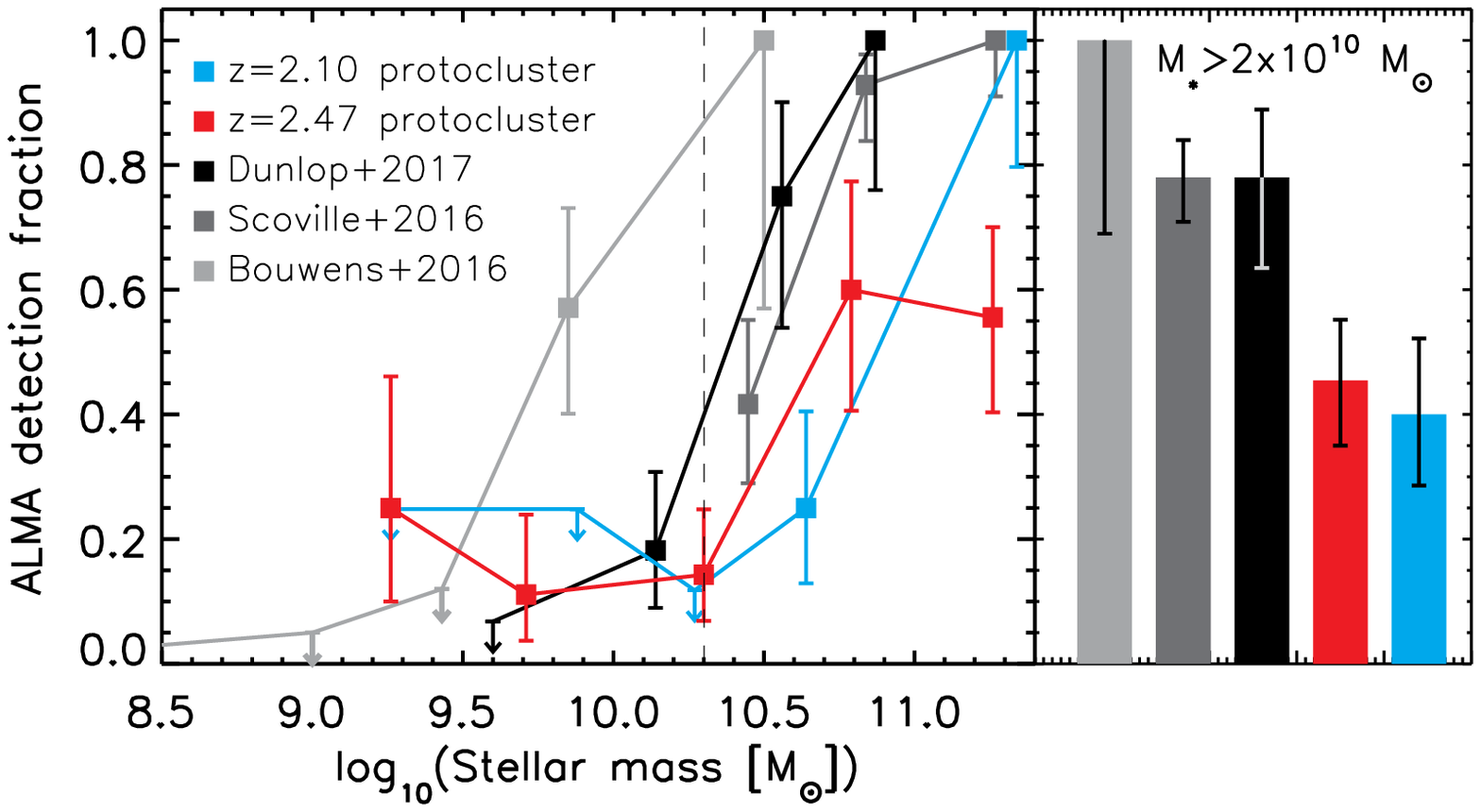}\vspace{0.7cm}
\caption{ALMA dust continuum detection fraction as a function of stellar mass for galaxies within $z\approx2-3$. Previous ALMA blank-field observations (\citealt{Bouwens2016a,Dunlop2017a}) and follow-ups of field galaxies (\citealt{Scoville2016a}) show a higher detection fraction than the one achieved towards the proto-cluster galaxies studied here (colored in red and blue), despite having similar depths. The difference is clearly evidenced by the detection fraction of the most massive galaxies ($M_\star>2\times10^{10}\,M_\odot$), as shown in the right panel. This implies a higher fraction of quenched gas-poor galaxies in the proto-cluster structures than the one found in the field at the same cosmic epoch, suggesting that massive galaxies in dense environments undergo an accelerated evolution. 
\label{fig:detection_fraction}}
\end{center}
\end{figure*}

In Figure \ref{fig:detection_fraction}, we compare the ALMA continuum detection fraction our proto-cluster galaxies to the one derived for our control sample (\citealt{Scoville2016a}) and also to those found in  {\it blind} blank-field observations (\citealt{Bouwens2016a,Dunlop2017a}), limiting these samples to  galaxies within $z\approx2-3$. Note that all the observations have similar depths\footnote{The average depth of $\sigma\approx150\rm\, \mu Jy$ at $870\rm\,\mu m$ of the observations used by \citealt{Scoville2016a} scales to $\sim50\rm\, \mu Jy$ at the wavelength used in this work, assuming the Rayleigh-Jeans relation and an emissivity index of $\beta=1.5$. Similarly, the depth of the map presented in \citet{Dunlop2017a} corresponds to $\sim45\rm\, \mu Jy$. These values are in good agreement with the depth of our observations (see \S\ref{secc:data}). The work by \citet{Bouwens2016a} is, however, based on observations a factor of $\sim3$x deeper.}, making the comparison appropriate. 

Interestingly, the observations towards proto-cluster member galaxies show a lower detection fraction than those for  the field samples, particularly for the high-mass galaxies (Figure \ref{fig:detection_fraction}). 
This is also true even if we only compare our results to those from the {\it blind} blank-field surveys, which have no biases due to selection effects. 
This difference is found in both the $z=2.10$ and the $z=2.47$ proto-cluster structures. These results  imply that {\it the most massive galaxies in proto-cluster environments experience an accelerated evolution} compared to the field sources, given the higher fraction of quiescent gas-poor galaxies.

We also estimate an average environmental quenching efficiency, defined as $\epsilon_q\equiv(f_{\rm pass,dense}-f_{\rm pass,field})/f_{\rm SFG,field}$, which represents the fraction of field galaxies required to be quenched in order to match the observed quenching fraction in the proto-clusters. For the two structures studied here, we estimate an average 
quenching efficiency of $\epsilon_q=0.45^{+0.16}_{-0.13}$. This quenching efficiency is indeed in agreement with those found for more evolved clusters at $z\approx1.0-1.6$ (see compilation by \citealt{Nantais2016a}). All these results suggest that we are witnessing the first stages of the  well-known environmental quenching found in lower redshift and/or more evolved clusters  (e.g. \citealt{Tran2009a, Alberts2014a,Newman2014a,Nantais2016a,Ji2018a,Lee-Brown2017a,Noirot2018a,Paulino-Afonso2018a,Shimakawa2018a}), which originates the excess of red sources in local galaxy clusters (\citealt{Dressler1984a}).

Understanding the mechanisms that drive this  evolution requires further observations than are included here. However, it is clear that these mechanisms should involve gas consumption or removal, or at least a halting of gas accretion, given the small gas fractions derived for these red galaxies (see also \citealt{Man2019a}. Note that counterexamples to this exist, i.e. passive galaxies with significant gas reservoirs (e.g. \citealt{Rowlands2015a,Suess2017a,Gobat2018a}), although in a variety of galaxy environments from isolated system to small groups.
Some of the typically cited quenching processes in high-density environments are AGN feedback (e.g. \citealt{Kauffmann2003a,Fabian2012a,Shimakawa2018a}), major and minor mergers (e.g. \citealt{Davis2018a,Maltby2018a}), ram-pressure stripping  (meaning gas being removed from the galaxies by the IGM, e.g. \citealt{Gunn1972a,Larson1980a,Boselli2006a,Foltz2018a}, although see \citealt{Dannerbauer2017a}), and starvation  (when gas accretion is stopped).

In Figure \ref{fig:UVcolors}, we identify those galaxies which are likely to host an AGN
based on their detection with {\it Chandra} X-ray Observatory (using the match reported by \citealt{Laigle2016a} to the {\it Chandra} COSMOS catalogs by \citealt{Civano2016a,Marchesi2016a}). Only a few galaxies are X-ray detected (all of them have $M_\star\gtrsim3\times10^{10}\,M_\odot$; see Tables \ref{table:catalogue1} and \ref{table:catalogue2}), and there is not a clear trend between AGN activity and quenching since some of them are detected by ALMA while others are not. Nevertheless, we cannot rule out this mechanism as the main quenching procces given the incompleteness related to the AGN identification. 

To shed light on other quenching mechanisms possibly involved, we estimate an upper limit on the quenching timescale adopting the average mass-weighted stellar age of the massive gas-poor galaxies derived from the SED fitting (see \S\ref{secc:sed_fitting}). Given that this quantity is related to the time since the onset of star formation\footnote{Note that the mass-weighted age estimated by {\sc MAGPHYS} depends not only on the time since the onset of star formation, but also on the shape of the star formation history. In the model, this quantity is typically $\sim1.2\times$ larger than the $R$-band light-weighted ages, the ages of the stars dominating the rest-frame R-band light (see details in \citealt{da-Cunha2015a}).}, any quenching mechanism must have a shorter timescale.  To be conservative, we adopt the longest age from the 97.5$^{th}$ percentile of the stellar age distribution, resulting in an upper limit on quenching timescale of $\tau_q<1\,\rm Gyr$. Similar timescales have been reported for passive galaxies in lower redshift clusters (e.g. \citealt{Muzzin2014a,Davis2018a,Socolovsky2018a}) and are usually associated with ram-pressure stripping and/or mergers (e.g. \citealt{Steinhauser2016a}).  In line with these findings, \citet{Casey2015a} and \citet{Hung2016a} found slightly higher fractions of irregular and interacting galaxies in the $z=2.47$ and 2.10 structures than in their control samples (although with a low statistical significance of $\sim1.5\sigma$), supporting the influence of galaxy mergers in the evolution of galaxies in overdense environments.

Given that the proto-cluster structures studied here are still in an assembly phase, these results indicate  galaxy pre-processing, meaning the interactions between groups of galaxies prior to the settling in a cluster potential, might be also an important quenching mechanism (see also \citealt{Zabludoff1996a,Haines2015a,Bianconi2018a,Olave-Rojas2018a}).

Alternatively, the advanced evolutionary stage of some massive proto-cluster member galaxies can be explained by an earlier onset of the star formation activity (e.g. \citealt{Steidel2005a}). Nevertheless, the mean stellar age of the red, massive gas-poor galaxies is consistent with the ages derived for the ALMA-detected galaxies (when using the same stellar mass threshold). Similarly, the stellar ages derived for the \citet{Scoville2016a} sample, representative of the average field population, 
show values in agreement with the proto-cluster galaxies studied here. Although we cannot rule out specific sources with premature star formation, we conclude that environmental effects, as those discussed above, most likely nurture an accelerated evolution in these systems.

% This value is in agreement with the rapid quenching scenarios (Muzzin2014a)
% 
% Muzzin2014afound that  quench on a rapid timescale (0.1 < τQ < 0.5 Gyr) for population of poststarburst galaxies in clusters at z ∼ 1 
% The derived quenching location and timescale suggest that gas stripping processes are most likely responsible for quenching the satellite population;
% 
% Steinhauser2016a, ram-pressure stripping  on timescale of ~200 Myr from simlations.
% 
% From simullations Davis2018a, quenching timescale of 600Myr for post-starburst galaxies from minor and major mergers and ram-pressuer stripping
% 
% Socolovsky2018a, supports rapid <1Gyr quenched procces (z=0.5-1)
% 
% pre-processing (e.g. Haines2015a, Olave-Rojas2018a)

\section{Conclusions} \label{secc:conclusions}

We have presented the largest census of molecular gas mass in proto-cluster member galaxies at $z\gtrsim2$, 
which combined with multiwavelength observations and SED modeling, allowed us to characterize the gas mass fraction and star formation efficiency for these sources, and to inquire into the environmental quenching during the earliest phases of clusters assembly. 

ALMA Band 6 ($\nu=233\,$GHz) observations were obtained for a total of 68 spectroscopically-confirmed galaxies within two large and massive overlapping proto-clusters in the COSMOS field at $z=2.10$ and 2.47, respectively, comprising most of the known massive galaxies with $M_\star\gtrsim10^{10}\,M_\odot$ in these structures, plus other less massive ($10^9\,M_\odot < M_\star < 10^{10}\,M_\odot$) galaxies which lie within the field of view. Given the low completeness level for the low-mass galaxies, our conclusions regard only to the most massive systems ($M_\star\gtrsim10^{10}\,M_\odot$).  

ISM masses were derived from the Rayleigh-Jeans dust continuum emission probed by the ALMA data, following the methodology developed by \citet{Scoville2014a,Scoville2016a}, while SFRs and stellar masses were estimated by fitting the photometric SED using {\sc MAGPHYS} (\citealt
{da-Cunha2008a,da-Cunha2015a}) and the rich multiwavelength photometry compiled by \citet{Laigle2016a} (see \S\ref{secc:observations}). All the derived properties are reported in Tables \ref{table:catalogue1}--\ref{table:catalogue3}.

Our analysis showed that, at the probed evolutionary stage of these systems, the star formation efficiency of most of the  proto-cluster members are similar to those found for coeval field galaxies and are in agreement with the field scaling relations (see Figure \ref{fig:SFE}), although, a non-negligible fraction of the least massive systems might have enhanced efficiencies. Most of these proto-cluster galaxies have also gas fractions that resemble those estimated for coeval field galaxies (Figure \ref{fig:gas_fract}), with the exception of a few of the most massive systems showing very low gas masses.

The effects of the environment are more evident when looking at the fraction of quenched galaxies. The larger number of massive gas-poor galaxies (ALMA non-detections) in the proto-clusters in comparison to the field  (Figures \ref{fig:gas_fract} and \ref{fig:detection_fraction}) suggests that these proto-cluster galaxies are undergoing an accelerated evolution (see also \citealt{Hayashi2018a,Shimakawa2018a,Wang2018a}).  These massive gas-poor systems have low gas mass fractions of $f_{\rm gas}\lesssim6-10\,\%$ (Figure \ref{fig:gas_fract}) and red colors (Figures \ref{fig:UVcolors} and \ref{fig:UVJ}) which are in agreement with those found for post-starburst and quiescent galaxies at lower redshifts (e.g. \citealt{Sargent2015a,Belli2018a,Gobat2018a}).

Environmental quenching is therefore manifested during the early phases of cluster assembly and must involve rapid mechanisms (on a timescale of hundreds of Myr). From our observations, we derive an upper limit on the quenching timescale of $\tau_q<1\,$Gyr  and a quenching efficiency of $\epsilon_q\equiv(f_{\rm passive,dense}-f_{\rm passive,field})/f_{\rm SFG,field} = 0.45^{+0.16}_{-0.13}$. Note that these estimations concern only galaxies with $M_\star\gtrsim2\times10^{10}\,M_\odot$ which are considered passive given their low gas content and red colors (see details in \S\ref{secc:quenching}).
Some of the processes typically cited in the literature with such timescales include AGN feedback, ram-pressure stripping, and galaxy mergers.
Regardless of their relatively importance, it is clear that these mechanisms should involve gas consumption or removal, or at least to halt gas accretion, given the small gas fractions derived for these red galaxies, nevertheless, further observations than those analyzed here are required to draw further conclusions.

Given that the proto-cluster studied here have not yet collapsed, our results suggest that quenching before virialization, also known in the literature as galaxy pre-processing, is an important mechanism related to the environmental quenching.

\acknowledgments

{\small {\it Acknowledgments:}
We thank Anthony Remijan and Jeremy Thorley from the North American ALMA Science Center (NAASC) for their help with data reduction and retrieval.
JAZ and CMC 
thank the University of Texas at Austin College of 
Natural Sciences for support, in addition to  NSF grant AST-1714528 and AST-1814034. ET acknowledges support from FONDECYT Regular 1160999, CONICYT PIA ACT172033 and Basal-CATA PFB-06/2007 and AFB170002 grants. HD acknowledges financial support from the Spanish Ministry of Science, Innovation and Universities (MICIU) under the 2014 Ramón y Cajal program RYC-2014-15686 and AYA2017-84061-P, the later one co-financed by FEDER (European Regional Development Funds). ST acknowledge support from the European Research Council (ERC) Consolidator Grant funding scheme (project ConTExt, grant No. 648179). The Cosmic Dawn Center is funded by the Danish National Research Foundation.
 
This paper makes use of the following ALMA data:
ADS/JAO.ALMA\#2016.1.00646.S. ALMA is a partnership of ESO (representing its member states), NSF (USA) and NINS (Japan), together with NRC (Canada) and NSC and ASIAA (Taiwan) and KASI (Republic of Korea), in cooperation with the Republic of Chile. The Joint ALMA Observatory is operated by ESO, AUI/NRAO and NAOJ.
}

% \facilities{ALMA}




\bibliography{paper_arxiv}
% \begin{thebibliography}{}
% \bibitem[Astropy Collaboration et al.(2013)]{2013A&A...558A..33A} Astropy Collaboration, Robitaille, T.~P., Tollerud, E.~J., et al.\ 2013, \aap, 558, A33 
% \bibitem[Bertin \& Arnouts(1996)]{1996A&AS..117..393B} Bertin, E., \& Arnouts, S.\ 1996, \aaps, 117, 393 
% \bibitem[Corrales(2015)]{2015ApJ...805...23C} Corrales, L.\ 2015, \apj, 805, 23
% \bibitem[Ferland et al.(2013)]{2013RMxAA..49..137F} Ferland, G.~J., Porter, R.~L., van Hoof, P.~A.~M., et al.\ 2013, \rmxaa, 49, 137
% \bibitem[Hanisch \& Biemesderfer(1989)]{1989BAAS...21..780H} Hanisch, R.~J., \& Biemesderfer, C.~D.\ 1989, \baas, 21, 780 
% \bibitem[Lamport(1994)]{lamport94} Lamport, L. 1994, LaTeX: A Document Preparation System, 2nd Edition (Boston, Addison-Wesley Professional)
% \bibitem[Schwarz et al.(2011)]{2011ApJS..197...31S} Schwarz, G.~J., Ness, J.-U., Osborne, J.~P., et al.\ 2011, \apjs, 197, 31  
% \bibitem[Vogt et al.(2014)]{2014ApJ...793..127V} Vogt, F.~P.~A., Dopita, M.~A., Kewley, L.~J., et al.\ 2014, \apj, 793, 127  
% 
% \end{thebibliography}

\appendix\label{appendix}

\section{Comparison to other scaling relations. }\label{appendix0} 
\begin{figure}[h!]\hspace{-0.3cm}
\includegraphics[width=0.53\textwidth]{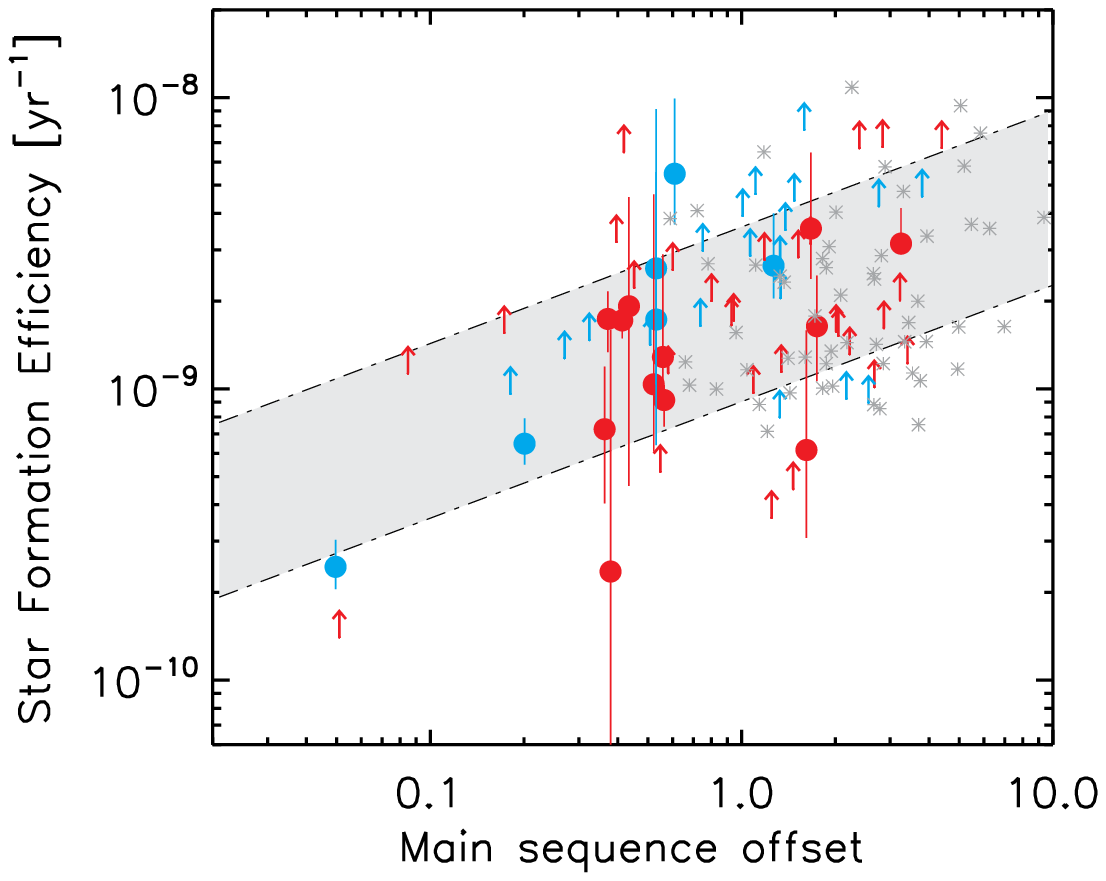}\hspace{-0.4cm}\includegraphics[width=0.53\textwidth]{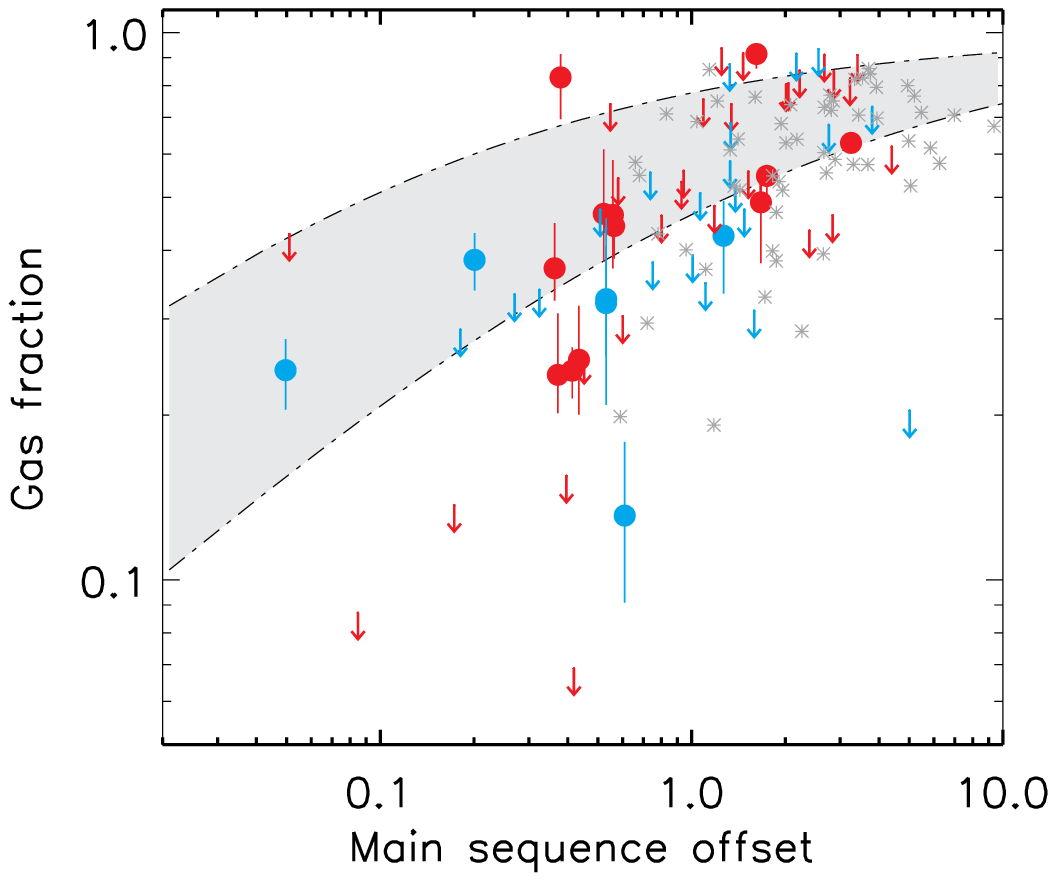}
\caption{{\it left:} Distribution of SFEs  (SFR/$M_{\rm gas}$; {\it left:}) and gas fractions ({\it right}) as a function of main sequence offset (SFR/$\rm SFR_{MS}$) for the proto-cluster member galaxies studied in this work. Those galaxies in the $z=2.47$ overdensity detected by ALMA are represented by the blue solid circles while non-detections are illustrated by the blue asterisks. Those in the $z=2.10$ are shown with red symbols (solid circles and asterisks for the detections and non-detections, respectively). Gray asterisks correspond the control sample adopted in our study (\citealt{Scoville2016a}). The gray shaded area represents the \citet{Tacconi2018a} field scaling relations for main-sequence galaxies at a redshift and stellar mass representative of our sample with $\pm0.3$\,dex scatter. 
\label{fig:SFE_msoffset}}
\end{figure}

The main results of this paper remain if we instead adopt the scaling relations derived by \citet{Tacconi2018a}, which take into account stellar masses and SFRs simultaneously. Figure \ref{fig:SFE_msoffset} shows the SFE as a function of offset from the main sequence ($\rm SFR/SFR_{MS}$, left panel) for the proto-cluster and field galaxies. Almost 80\% of the detected sources (14 out of 18) lie on the \citeauthor{Tacconi2018a} scaling relation. Interestingly, there is a non-negligible fraction of galaxies whose lower limits on the SFEs place them above the field relation. However,
most of these sources have stellar masses below $M_*\sim10^{10}\,M_\odot$, probing a stellar mass range that suffers from low completeness in our sample, which prevents us from investigating the relevance of these galaxies in comparison to the bulk of the population. As mentioned above, the conclusions from this work focus only on the most massive objects with $M_*\gtrsim10^{10}\,M_\odot$. On the other hand, most of the gas fractions of the proto-cluster member galaxies (right panel) lie on or below the field relation. These results suggest that proto-cluster galaxies undergo an accelerated evolution, which produces an enhanced fraction of passive gas-poor galaxies than relative to the field.

\section{Derived properties}\label{appendix1}

\startlongtable
\begin{deluxetable*}{lcccccc}
\tablecaption{Properties of the $z=2.10$ proto-cluster galaxies observed with ALMA.  \label{table:catalogue1}}
\tablecolumns{9}
\tablenum{1}
\tablewidth{0pt}
\tablehead{
\colhead{ID} &
\colhead{$z_{\rm spec}$} &
\colhead{SNR} &
\colhead{$S_{1.3{\rm\,mm}}$} &
\colhead{$M_\star$\tablenotemark{a}} &
\colhead{SFR\tablenotemark{a}} &
\colhead{$M_{\rm ISM}$\tablenotemark{b}}  \\
\colhead{} & 
\colhead{} & 
\colhead{} & 
\colhead{[$\rm \mu Jy$]} & 
\colhead{[$\times10^{10}\,M_\odot$]} & 
\colhead{[$\sfr$]} & 
\colhead{[$\times10^{10}\,M_\odot$]} 
}
\startdata
% ID1\tablenotemark{$b$} & XX:XX:XX.X & $-$XX:XX:XX.X &  2.10  & 3.5 & $XX\pm XX$ & $XX\pm XX$ & $XX\pm XX$ &  $XX\pm XX$\tablenotemark{$c$}         \\
 J100031.84$+$021242.7   & 2.1043 & $ 51 $    & $2301\pm  45 $    & $ 30.9_{-0.7 }^{+6.3 } $ & $ 612_{-200}^{+280} $ & $  37.3 \pm  0.7 $ \\   %z21_39   
 J100041.25$+$021426.5   & 2.0880 & $21.8 $    & $954 \pm  43 $    & $ 26.3_{-6.4 }^{+4.6 } $ & $ 113_{-47}^{+64}   $ & $  15.5 \pm  0.7 $ \\  %z21_10   
 J100032.73$+$021331.1\tablenotemark{c}   & 2.0908 & $16.4 $    & $704 \pm  42 $    & $  6.7_{-0.0 }^{+0.3 } $ & $ 361_{-8}^{+88}    $ & $  11.4 \pm  0.6 $ \\  %z21_94   
 J100018.24$+$021242.5   & 2.1021 & $ 8.9 $    & $892 \pm  100$    & $ 18.1_{-0.0 }^{+2.2 } $ & $ 132_{-13}^{+0}    $ & $  14.4 \pm  1.6 $ \\  %z21_114  
 J100042.65$+$020850.9   & 2.0988 & $ 7.2 $    & $324 \pm  45 $    & $ 16.5_{-0.0 }^{+0.0 } $ & $ 91_{-1}^{+0}      $ & $   5.2 \pm  0.7 $ \\  %z21_104  
 J100046.65$+$021623.9   & 2.0981 & $ 6.0 $    & $208 \pm  34 $    & $ 10.0_{-1.5 }^{+1.2 } $ & $ 65_{-47}^{+63}    $ & $   3.3 \pm  0.5 $ \\  %z21_44   
 J095936.45$+$021614.5\tablenotemark{c}   & 2.0903 & $ 5.4 $    & $214 \pm  39 $    & $ 11.2_{-1.9 }^{+0.0 } $ & $ 60_{-5}^{+1}      $ & $   3.4 \pm  0.6 $ \\  %z21_62   
 J100016.01$+$021527.8   & 2.1019 & $ 5.2 $    & $289 \pm  55 $    & $  5.3_{-1.8 }^{+0.8 } $ & $ 50_{-15}^{+130}   $ & $   4.6 \pm  0.8 $ \\  %z21_43   
 J100000.91$+$020902.6   & 2.0986 & $ 4.0 $    & $162 \pm  41 $    & $  2.7_{-0.1 }^{+0.5 } $ & $ 94_{-15}^{+34}    $ & $   2.6 \pm  0.6 $ \\  %z21_76   
 J100015.02$+$021538.9   & 2.0922 & $ 3.2 $    & $167 \pm  51 $    &   $0.5_{-0.2 }^{+0.3 } $ & $ 6_{-5}^{+23}      $ & $   2.7 \pm  0.8 $ \\  %z21_20   
 J100022.64$+$021434.6   & 2.0979 & $ 3.2 $    & $106 \pm  32 $    & $  0.1_{-0.04}^{+0.03} $ & $ 11_{-4}^{+8}      $ & $   1.7 \pm  0.5 $ \\  %z21_11   
 J100018.06$+$021409.9   & 2.0951 & $ 3.1 $    & $161 \pm  51 $    & $  3.0_{-0.6 }^{+0.0 } $ & $ 34_{-3}^{+18}     $ & $   2.6 \pm  0.8 $ \\  %z21_53   
 J100018.55$+$021817.2   & 2.0924 & $<3.0 $    & $<236        $    & $  1.3_{-0.2 }^{+0.4 } $ & $ 44_{-23}^{+46}    $ & $  <3.8          $ \\  %z21_4    
 J095949.58$+$022445.3   & 2.0850 & $<3.0 $    & $<120        $    & $  1.6_{-0.5 }^{+1.0 } $ & $ 22_{-4}^{+13}     $ & $  <1.9          $ \\  %z21_66   
 J100018.60$+$021257.7   & 2.0859 & $<3.0 $    & $<139        $    & $  2.6_{-0.6 }^{+0.0 } $ & $ 154_{-45}^{+0}    $ & $  <2.2          $ \\  %z21_109  
 J100040.82$+$021822.9   & 2.1001 & $<3.0 $    & $<120        $    & $  1.5_{-0.3 }^{+0.8 } $ & $ 55_{-17}^{+88}    $ & $  <1.9          $ \\  %z21_100  
 J100022.73$+$021423.6   & 2.1027 & $<3.0 $    & $<194        $    & $  2.7_{-0.3 }^{+0.2 } $ & $ 52_{-20}^{+19}    $ & $  <3.1          $ \\  %z21_57   
 J100015.74$+$021539.6   & 2.1040 & $<3.0 $    & $<114        $    & $ 19.4_{-0.0 }^{+3.4 } $ & $ 21_{-0}^{+14}     $ & $  <1.8          $ \\  %z21_40   
 J100023.69$+$021604.1   & 2.0920 & $<3.0 $    & $<121        $    & $  0.1_{-0.1 }^{+0.0 } $ & $ 20_{-12}^{+0}     $ & $  <1.9          $ \\  %z21_28   
 J100028.14$+$021325.9   & 2.0857 & $<3.0 $    & $<117        $    & $ 12.0_{-0.3 }^{+0.0 } $ & $ 30_{-1}^{+1}      $ & $  <1.9          $ \\  %z21_49   
 J100022.53$+$021556.3   & 2.0978 & $<3.0 $    & $<229        $    & $  0.3_{-0.15}^{+0.18} $ & $ 17_{-10}^{+21}    $ & $  <3.7          $ \\  %z21_116  
 J100015.56$+$022029.6   & 2.0900 & $<3.0 $    & $<126        $    & $  1.6_{-0.1 }^{+0.2 } $ & $ 65_{-27}^{+34}    $ & $  <2.0          $ \\  %z21_82   
 J100029.98$+$021413.1   & 2.0985 & $<3.0 $    & $<109        $    & $  4.0_{-0.0 }^{+0.0 } $ & $ 45_{-1}^{+1}      $ & $  <1.7          $ \\  %z21_55   
 J100023.02$+$021434.4   & 2.0961 & $<3.0 $    & $<116        $    & $  0.3_{-0.18}^{+0.08} $ & $ 25_{-15}^{+12}    $ & $  <1.8          $ \\  %z21_90   
 J100022.33$+$021441.7   & 2.0986 & $<3.0 $    & $<154        $    & $  0.8_{-0.25}^{+0.23} $ & $ 24_{-13}^{+16}    $ & $  <2.5          $ \\  %z21_12   
 J100025.04$+$021005.9   & 2.0962 & $<3.0 $    & $<103        $    & $  2.1_{-0.1 }^{+1.0 } $ & $ 113_{-57}^{+28}   $ & $  <1.6          $ \\  %z21_92   
 J100023.61$+$021557.3   & 2.0893 & $<3.0 $    & $<152        $    & $  0.4_{-0.2 }^{+0.1 } $ & $ 40_{-30}^{+14}    $ & $  <2.4          $ \\  %z21_25   
 J100017.90$+$021807.1\tablenotemark{c}   & 2.0937 & $<3.0 $    & $<134        $    & $ 29.5_{-3.8 }^{+6.0 } $ & $ 140_{-40}^{+240} $ & $  <2.1          $ \\   %z21_38   
 J100022.99$+$021605.2   & 2.0943 & $<3.0 $    & $<124        $    & $  0.1_{-0.07}^{+0.09} $ & $ 7_{-5}^{+10}      $ & $  <2.0          $ \\  %z21_5    
 J100044.05$+$021522.2   & 2.0914 & $<3.0 $    & $<119        $    & $  2.0_{-0.1 }^{+0.0 } $ & $ 54_{-1}^{+1}      $ & $  <1.9          $ \\  %z21_17   
 J100022.19$+$021306.3   & 2.0978 & $<3.0 $    & $<121        $    & $  2.6_{-0.8 }^{+0.0 } $ & $ 3_{-0}^{+6}       $ & $  <1.9          $ \\  %z21_108  
 J095955.88$+$022459.0   & 2.0899 & $<3.0 $    & $<115        $    & $  2.1_{-0.4 }^{+0.3 } $ & $ 38_{-23}^{+27}    $ & $  <1.8          $ \\  %z21_71   
 J100019.19$+$021406.5   & 2.1014 & $<3.0 $    & $<127        $    & $  6.0_{-1.2 }^{+0.7 } $ & $ 46_{-18}^{+35}    $ & $  <2.0          $ \\  %z21_52   
 J100015.89$+$021543.7   & 2.1090 & $<3.0 $    & $<129        $    & $  1.6_{-0.3 }^{+0.5 } $ & $ 36_{-20}^{+13}    $ & $  <2.1          $ \\  %z21_21   
 J100015.89$+$021547.2   & 2.0976 & $<3.0 $    & $<151        $    & $  0.8_{-0.23}^{+0.3 } $ & $ 13_{-7}^{+20}     $ & $  <2.4          $ \\  %z21_24   
 J095950.32$+$022206.0   & 2.0902 & $<3.0 $    & $<122        $    & $  1.2_{-0.2 }^{+0.3 } $ & $ 134_{-83}^{+2}    $ & $  <1.9          $ \\  %z21_67   
 J100046.37$+$021622.0   & 2.0891 & $<3.0 $    & $<118        $    & $ 10.4_{-1.4 }^{+1.6 } $ & $ 61_{-24}^{+24}    $ & $  <1.9          $ \\  %z21_33   
 J100023.30$+$021612.6   & 2.0950 & $<3.0 $    & $<161        $    & $  0.2_{-0.06}^{+0.12} $ & $ 32_{-24}^{+0}     $ & $  <2.6          $ \\  %z21_32   
 J100023.38$+$021606.3   & 2.0917 & $<3.0 $    & $<122        $    & $  0.4_{-0.0 }^{+0.1 } $ & $ 30_{-0.00}^{+16}  $ & $  <1.9          $ \\  %z21_29   
 J100018.53$+$021306.9   & 2.1056 & $<3.0 $    & $<179        $    & $  0.6_{-0.04}^{+0.36} $ & $ 58_{-28}^{+40}    $ & $  <2.9          $ \\  %z21_113  
 J100022.86$+$021432.3   & 2.0973 & $<3.0 $    & $<107        $    & $  0.4_{-0.12}^{+0.12} $ & $ 27_{-15}^{+15}    $ & $  <1.7          $\\   %z21_58   
\enddata
\tablecomments{$^a$The SFRs and stellar masses were derived through an SED-fitting precedure (see \S\ref{secc:sed_fitting}).
$^b$ISM masses were estimated from dust-continuum emission (see \S\ref{secc:ism_mass}). $^c$ X-ray detected sources. All the reported upper limits are $3\sigma$ values.
}
\end{deluxetable*}

\begin{deluxetable*}{lcccccc}[h]
\tablecaption{Properties of the $z=2.47$ proto-cluster galaxies observed with ALMA.  \label{table:catalogue2}}
\tablecolumns{9}
\tablenum{2}
\tablewidth{0pt}
\tablehead{
\colhead{ID} &
\colhead{$z_{\rm spec}$} &
\colhead{SNR} &
\colhead{$S_{1.3{\rm\,mm}}$} &
\colhead{$M_\star$\tablenotemark{a}} &
\colhead{SFR\tablenotemark{a}} &
\colhead{$M_{\rm ISM}$\tablenotemark{b}}  \\
\colhead{} & 
\colhead{} & 
\colhead{} & 
\colhead{[$\rm \mu Jy$]} & 
\colhead{[$\times10^{10}\,M_\odot$]} & 
\colhead{[$\sfr$]} & 
\colhead{[$\times10^{10}\,M_\odot$]}
}
\startdata
% ID1\tablenotemark{$b$} & XX:XX:XX.X & $-$XX:XX:XX.X &  2.10  & 3.5 & $XX\pm XX$ & $XX\pm XX$ & $XX\pm XX$ &  $XX\pm XX$\tablenotemark{$c$}         \\
 J100056.95$+$022017.2   & 2.4940 & $44.7 $    & $ 2095 \pm 47$    & $     -                $ & $ -                 $ & $  32.3 \pm 0.7  $ \\  %450.28.1
 J100057.56$+$022011.1   & 2.5130 & $11.3 $    & $ 957 \pm 84 $    & $ 19.9_{-3.4 }^{+7.0 } $ & $ 390_{-60}^{+150}  $ & $  14.7 \pm 1.3  $ \\  %450.28.2
 J100057.26$+$022012.4   & 2.5040 & $10.0 $    & $ 583 \pm 58 $    & $ 18.6_{-4.2 }^{+4.8 } $ & $ 160_{-90}^{+290} $ & $   8.9 \pm 0.8  $ \\   %450.28.3
 J100056.85$+$022008.8   & 2.5030 & $ 6.9 $    & $ 347 \pm 51 $    & $ 35.4_{-7.3 }^{+10.2} $ & $ 290_{-70}^{+160} $ & $   5.3 \pm 0.7  $ \\   %450.28.4
 J100115.18$+$022349.7   & 2.4710 & $ 5.5 $    & $ 254 \pm 46 $    & $  6.3_{-0.1 }^{+0.0 } $ & $ 26_{-1}^{+1}      $ & $   3.9 \pm 0.7  $ \\  %oz28    
 J100057.38$+$022010.5   & 2.5080 & $ 5.2 $    & $ 384 \pm 74 $    & $ 18.6_{-0.0 }^{+0.0 } $ & $ 15_{-1}^{+1}      $ & $   5.9 \pm 1.1  $ \\  %450.28.5
 J100025.28$+$022643.3   & 2.4750 & $ 3.1 $    & $ 108 \pm 35 $    & $  3.5_{-0.9 }^{+0.7 } $ & $ 43_{-20}^{+59}    $ & $   1.6 \pm 0.5  $ \\  %ZCOS14  
 J100111.03$+$022043.3   & 2.4670 & $<3.0 $    & $<103         $   & $  1.5_{-0.6 }^{+0.3 } $ & $ 46_{-21}^{+52}    $ & $  <1.6          $ \\  %oz27    
 J100039.40$+$022155.4   & 2.4571 & $<3.0 $    & $<101         $   & $  0.7_{-0.2 }^{+0.1 } $ & $ 32_{-11}^{+55}    $ & $  <1.5          $ \\  %vuds7   
 J100020.50$+$022421.4   & 2.4720 & $<3.0 $    & $<101         $   & $  1.7_{-0.7 }^{+0.0 } $ & $ 70_{-1}^{+36}     $ & $  <1.5          $ \\  %ZCOS08  
 J100031.13$+$023103.3   & 2.4680 & $<3.0 $    & $<124         $   & $  0.1_{-0.05}^{+0.30} $ & $ 18_{-8}^{+32}     $ & $  <1.9          $ \\  %COMP422 
 J100015.38$+$022448.2   & 2.4740 & $<3.0 $    & $<104         $   & $  3.2_{-0.0 }^{+0.0 } $ & $ 21_{-1}^{+1}      $ & $  <1.6          $ \\  %ZCOS03  
 J100025.09$+$022500.3   & 2.4709 & $<3.0 $    & $<110         $   & $  0.8_{-0.1 }^{+0.2 } $ & $ 72_{-38}^{+17}    $ & $  <1.6          $ \\  %vuds13  
 J100013.61$+$022604.8   & 2.4630 & $<3.0 $    & $<102         $   & $  2.4_{-0.7 }^{+0.3 } $ & $ 62_{-32}^{+65}    $ & $  <1.5          $ \\  %ZCOS01  
 J100015.86$+$021939.5   & 2.4750 & $<3.0 $    & $<127         $   & $  4.3_{-1.5 }^{+0.6 } $ & $ 152_{-48}^{+60}   $ & $  <1.9          $ \\  %ZCOS04  
 J100059.45$+$021957.4\tablenotemark{c}   & 2.4710 & $<3.0 $    & $<127         $   & $  4.8_{-0.6 }^{+0.2 } $ & $ 19_{-3}^{+9}      $ & $  <1.9          $ \\  %COMP971 
 J100012.36$+$023707.5   & 2.4750 & $<3.0 $    & $<134         $   & $  3.8_{-0.8 }^{+0.5 } $ & $ 100_{-49}^{+130.} $ & $  <2.0          $ \\  %lz34    
 J100024.21$+$022741.3   & 2.4790 & $<3.0 $    & $<104         $   & $  1.4_{-0.2 }^{+0.1 } $ & $ 56_{-27}^{+22}    $ & $  <1.6          $ \\  %ZCOS12  
 J100008.88$+$023044.0   & 2.4750 & $<3.0 $    & $<140         $   & $  2.3_{-0.2 }^{+0.6 } $ & $ 31_{-9}^{+6}      $ & $  <2.1          $ \\  %lz33    
 J100027.12$+$023253.8   & 2.4745 & $<3.0 $    & $<121         $   & $  0.1_{-0.0 }^{+0.0 } $ & $ 16_{-1}^{+0}      $ & $  <1.8          $ \\  %vuds4   
 J100033.19$+$022225.0   & 2.4740 & $<3.0 $    & $<112         $   & $  0.2_{-0.06}^{+0.18} $ & $ 14_{-10}^{+10}    $ & $  <1.7          $ \\  %ZCOS00  
 J100109.29$+$022221.5   & 2.4730 & $<3.0 $    & $<124         $   & $  1.5_{-0.3 }^{+0.4 } $ & $ 32_{-11}^{+42}    $ & $  <1.9          $ \\  %oz26    
 J100018.03$+$021808.5   & 2.4720 & $<3.0 $    & $<134         $   & $  0.7_{-0.0 }^{+0.0 } $ & $ 95_{-1}^{+1}      $ & $  <2.0          $ \\  %ZCOS05  
 J100050.73$+$021922.4   & 2.4660 & $<3.0 $    & $<109         $   & $  2.7_{-0.5 }^{+0.2 } $ & $ 50_{-15}^{+28}    $ & $  <1.6          $ \\  %ZCOS19  
 J100014.23$+$022516.7   & 2.4710 & $<3.0 $    & $<127         $   & $  1.4_{-0.3 }^{+0.4 } $ & $ 53_{-24}^{+54}    $ & $  <1.9          $ \\  %ZCOS02  
 J100054.06$+$022104.3\tablenotemark{c}   & 2.4780 & $<3.0 $    & $<112         $   & $  3.3_{-0.1 }^{+0.4 } $ & $ 26_{-8}^{+9}      $ & $  <1.7          $ \\  %ZCOS21  
 J100021.97$+$022356.5\tablenotemark{c}   & 2.4730 & $<3.0 $    & $<102         $   & $  6.1_{-0.0 }^{+1.4 } $ & $ 630_{-250.}^{+0}  $ & $  <1.5          $ \\  %ZCOS09  
\enddata
\tablecomments{$^a$The SFRs and stellar masses were derived through an SED-fitting precedure (see \S\ref{secc:sed_fitting}).
$^b$ISM masses were estimated from dust-continuum emission (see \S\ref{secc:ism_mass}). $^c$ X-ray detected sources. All the reported upper limits are $3\sigma$ values.
}
\end{deluxetable*}

\begin{deluxetable*}{lccccc}[!h]
\tablecaption{Properties of the stacked subsamples.  \label{table:catalogue3}}
\tablecolumns{6}
\tablenum{3}
\tablewidth{0pt}
\tablehead{
\colhead{Subsample} &
\colhead{N} &
\colhead{$\langle z \rangle$} &
\colhead{$\langle M_\star \rangle$\tablenotemark{a}} &
\colhead{$\rm \langle SFR \rangle$\tablenotemark{a}} &
\colhead{$\langle M_{\rm ISM} \rangle$\tablenotemark{b}} \\ 
\colhead{} & 
\colhead{} &
\colhead{} & 
\colhead{[$\,M_\odot$]} & 
\colhead{[$\sfr$]} & 
\colhead{[$\,M_\odot$]}
}
\startdata
Galaxies with $10^9< M_\star<10^{10}\,M_\odot$ in the $z=2.10$ protocluster   & 11  & 2.0959 &  $3.8_{-2.4}^{+4.8}\times10^9$    & $23_{-15}^{+35}$ & $<7.6\times10^9$ \\
Galaxies with $10^{10}< M_\star<10^{11}\,M_\odot$ in the $z=2.10$ protocluster& 14  & 2.0950 &  $2.2_{-1}^{+3.8}\times10^{10}$   & $45_{-42}^{+109}$ & $9.0\pm1.9\times10^9$ \\
Galaxies with $M_\star>10^{11}\,M_\odot$ in the $z=2.10$ protocluster         & 4  & 2.0931 &  $1.6_{-0.5}^{+1.4}\times10^{11}$ & $23_{-15}^{+35}$ & $<1.0\times10^{10}$ \\
Galaxies with $10^9< M_\star<10^{10}\,M_\odot$ in the $z=2.47$ protocluster   & 6  & 2.4694 &  $3.7_{-2.4}^{+4.4}\times10^9$    & $31_{-17}^{+64}$ & $8.2\pm2.5\times10^{9}$ \\
Galaxies with $10^{10}< M_\star<10^{11}\,M_\odot$ in the $z=2.47$ protocluster& 14  & 2.4723 &  $2.6_{-1.2}^{+3.5}\times10^{10}$ & $55_{-36}^{+570}$ & $<5.0\times10^9$ \\
Galaxies with $10<{\rm SFR<100}\,\sfr$ in the $z=2.10$ protocluste        & 18  & 2.0957 &  $1.4_{-1.2}^{+18}\times10^{10}$  & $33_{-20}^{+30}$ & $<4.3\times10^9$ \\
Galaxies with $10<{\rm SFR<100}\,\sfr$ in the $z=2.47$ protocluster       & 23  & 2.4711 &  $1.3_{-1.1}^{+3.6}\times10^{10}$  & $37_{-23}^{+59}$ & $4.8\pm1.4\times10^{9}$  
\enddata
\tablecomments{$^a$The quoted uncertainties on the SFRs and stellar masses actually represent the width of the bins.
$^b$The stacked ISM masses were estimated from the upper limits on the dust continuum using an inverse variance weighting. All the reported upper limits are $3\sigma$ values.
}
\end{deluxetable*}

\newpage
\section{HST/ALMA cutouts}\label{appendix2}

\begin{figure*}[!h]
\includegraphics[width=\textwidth]{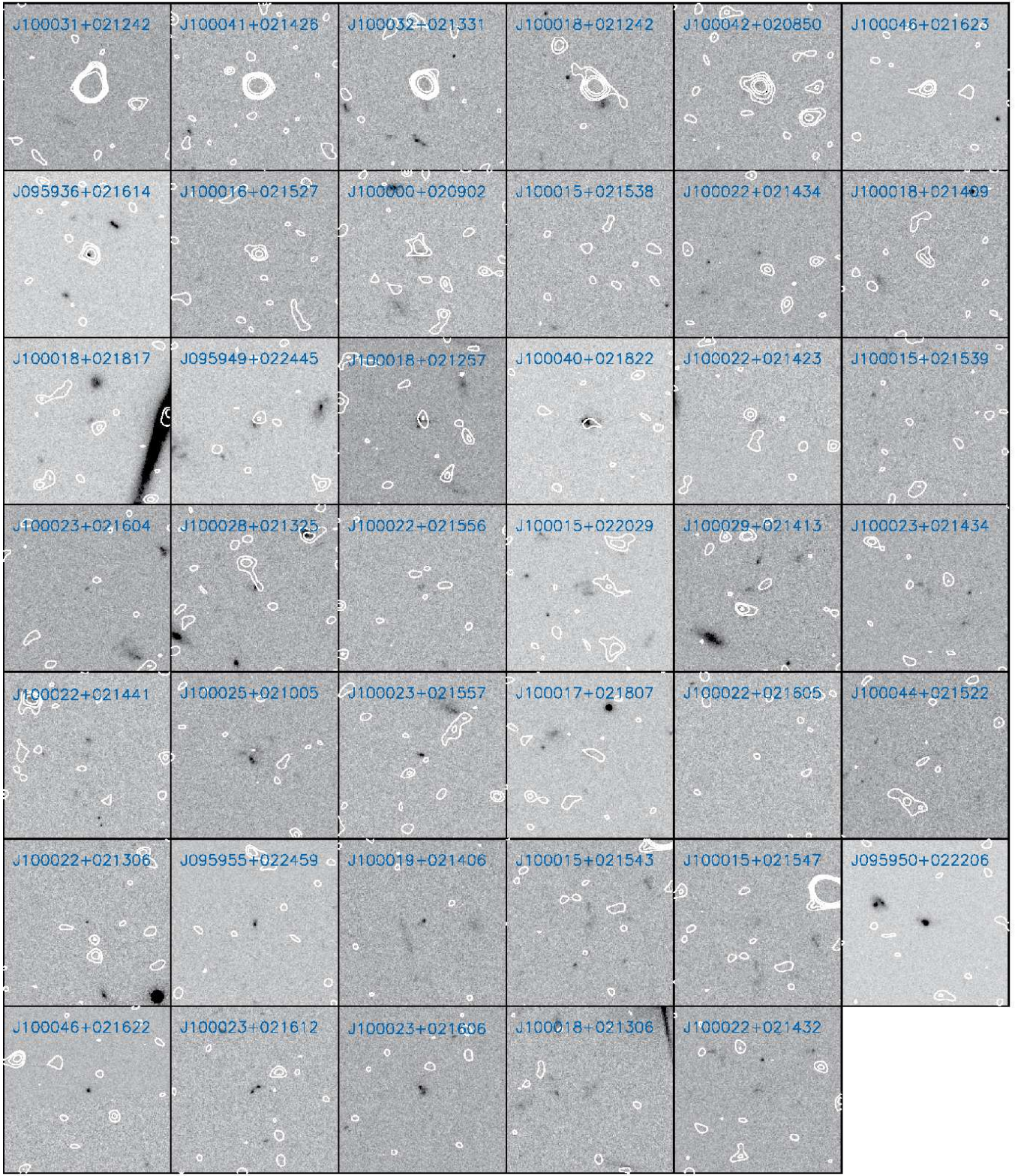}
\caption{10\,arcsec$\times$10\,arcsec postage stamps in the HST/ACS $I$-band for all the targets in the $z=2.10$ proto-cluster observed with ALMA. All the cutouts are centered at the optical position of each galaxy. The white contours show the 2, 3, 4.5, 6, and 10$\sigma$ of ALMA Band 6 ($\nu=233\rm\,GHz$, $\lambda\approx1.3\rm\,mm$) observations. The ID of each source is on the top of each panel.
\label{postage_stamps_1}}
\end{figure*}

\begin{figure*}[!h]
\includegraphics[width=\textwidth]{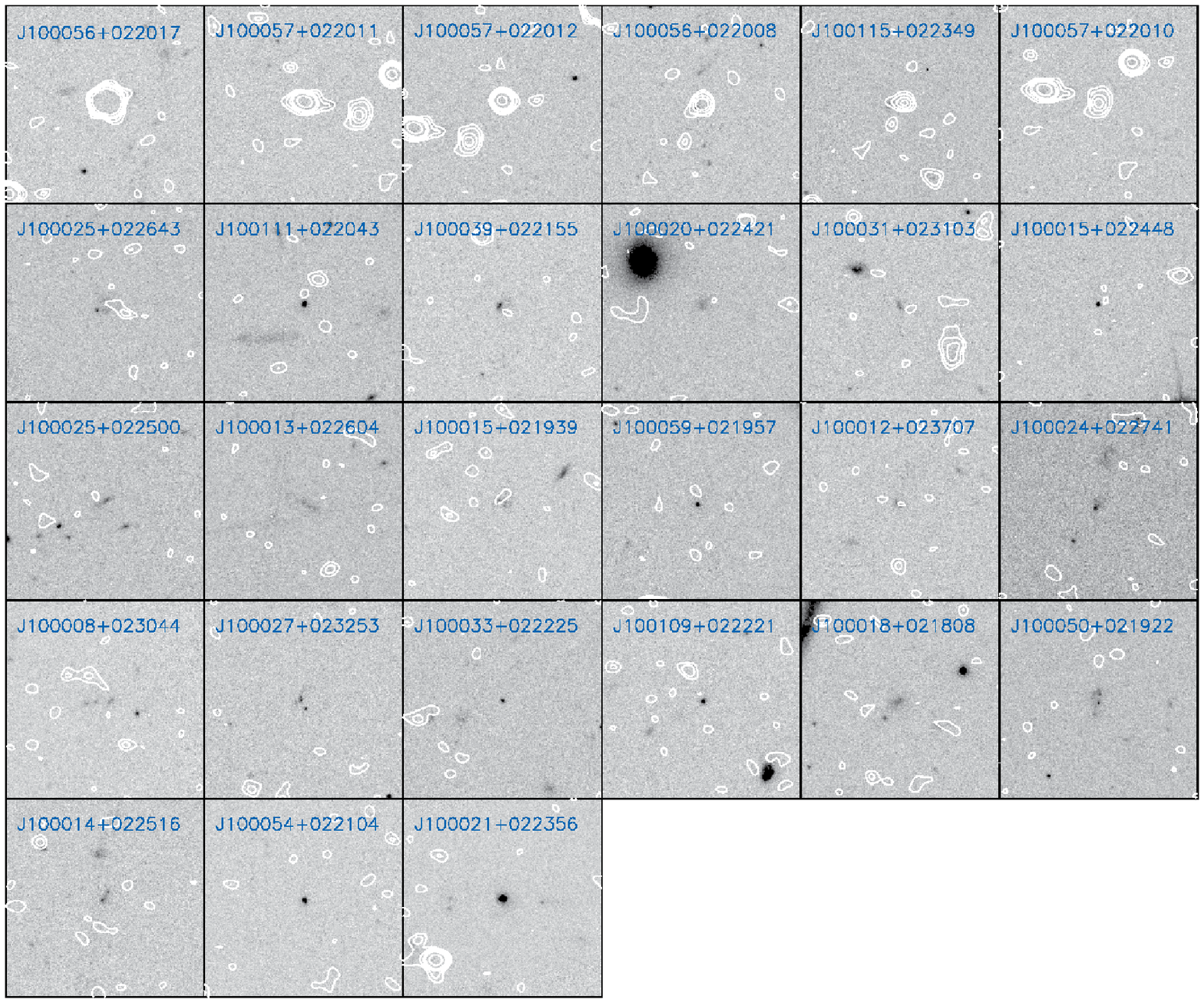}
\caption{10\,arcsec$\times$10\,arcsec postage stamps in the HST/ACS $I$-band for all the targets in the $z=2.47$ proto-cluster observed with ALMA.  All the cutouts are centered at the optical position of each galaxy. The white contours show the 2, 3, 4.5, 6, and 10$\sigma$ of ALMA Band 6 ($\nu=233\rm\,GHz$, $\lambda\approx1.3\rm\,mm$) observations. The ID of each source is on the top of each panel.
\label{postage_stamps_2}}
\end{figure*}



\end{document}